\documentclass[a4paper,12pt]{article}

\pdfoutput=1 

\usepackage[hmargin=.71in,vmargin=1.1in]{geometry}
\usepackage{indentfirst}
\usepackage{amsfonts}
\usepackage{mathrsfs}
\usepackage{amsmath}
\usepackage{mathabx}
\usepackage{amssymb}
\usepackage{authblk}
\usepackage{cite}
\usepackage{xcolor}
\usepackage{mathtools}
\usepackage{tensor}
\usepackage{graphicx}
\usepackage{bm}
\usepackage{upgreek}
\usepackage{braket}
\usepackage{color,soul}
\usepackage{csquotes}
\usepackage{caption}
\usepackage{subcaption}
\usepackage{slashed}
\usepackage{mathtools}
\usepackage{soul}

\usepackage[bookmarksnumbered=true,bookmarksopen=true]{hyperref}
 \hypersetup{colorlinks,
             linkcolor=[rgb]{0,0.3,0.6}, 
             citecolor=[rgb]{0,0.3,0.6}, 
             urlcolor=[rgb]{0,0.3,0.6}}

\newcommand\mi{\mathrm{i}}
\newcommand\me{\mathrm{e}}

\newcommand\pp{\uppi}
\newcommand{\dif}{\mathrm{d}}

\DeclareMathOperator{\arctanh}{arctanh}

\usepackage{tikz}
\usetikzlibrary{matrix}

\begin{document}

\title{\Large\textbf{Finite Curvature Construction of Regular Black Holes and Quasinormal Mode Analysis}}

\author[a]{Chen Lan\thanks{stlanchen@126.com}}
\author[b]{Zhen-Xiao Zhang \thanks{zx.zhang@mail.nankai.edu.cn}}
\author[c]{Hao Yang\thanks{Corresponding author: hyang@ucas.ac.cn}}

\affil[a]{\normalsize{\em Department of Physics, Yantai University, 30 Qingquan Road, Yantai 264005, China}}
\affil[b]{\normalsize{\em School of Physics, Nankai University, 94 Weijin Road, Tianjin 300071, China}}
\affil[c]{\normalsize{\em School of Fundamental Physics and Mathematical Sciences, Hangzhou Institute for Advanced Study, UCAS, Hangzhou 310024, China}}

\date{ }

\maketitle

\begin{abstract}
We develop a class of regular black holes by prescribing finite curvature invariants and reconstructing the corresponding spacetime geometry. Two distinct approaches are employed: one based on the Ricci scalar and the other on the Weyl scalar. In each case, we explore a variety of analytic profiles for the curvature functions, including Gaussian, hyperbolic secant, and rational forms, ensuring regularity, asymptotic flatness, and compatibility with dominant energy conditions. The resulting mass functions yield spacetime geometries free from curvature singularities and exhibit horizons depending on model parameters. 
To assess the stability of these solutions, we perform a detailed analysis of quasinormal modes (QNMs) under axial gravitational perturbations. 
We show that the shape of the effective potential, particularly its width and the presence of potential valleys, plays a critical role in determining the QNMs. 
Models with a large peak-to-valley ratio in the potential barrier tend to support longer-lived oscillations, since perturbations can be partially trapped in the valley region before eventually escaping.
By contrast, when the ratio is small, the valley is too shallow to produce effective trapping, and the waveforms reduce to standard exponential decay without sustained oscillatory behavior.
Our results highlight the significance of potential design in constructing physically viable and dynamically stable regular black holes, offering potential observational implications in modified gravity and quantum gravity scenarios.

\end{abstract}

\tableofcontents

\section{Introduction}
\label{sec:intr}
Spacetime singularities\cite{Hawking:1973uf,Wald:1984rg}, where curvature invariants diverge and classical general relativity fails, remain a profound challenge in gravitational physics.
Notably, while the Schwarzschild and Kerr spacetimes have been remarkably successful in modeling astrophysical black holes, their inherent central singularities lead to geodesic incompleteness, which is a fundamental limitation in these spacetime descriptions.
Therefore, since the seminal work of Bardeen\cite{Bardeen:1968nsg}, investigations of \textit{regular black holes}, that is, black holes free of curvature singularities \cite{Dymnikova:1992ux,Borde:1996df,Ayon-Beato:1998hmi,Bronnikov:2000vy}, have attracted considerable attention in both classical and quantum gravity frameworks\cite{Ansoldi:2008jw,Bambi:2023try,Lan:2023cvz}.

In general, regular black holes are constructed by modifying the mass function in the metric to ensure that curvature invariants remain finite throughout the spacetime. 
Such constructions can be motivated by various theoretical approaches, including nonlinear electrodynamics~\cite{Ayon-Beato:2000mjt,Capozziello:2024ucm}, loop quantum gravity~\cite{Modesto:2004xx}, and phenomenological geometric modifications~\cite{Fan:2016hvf,Maeda:2021jdc}. 
For further research on regular black holes, it is necessary to find systematic guiding principles and corresponding generalization methods without ad hoc assumptions.

In general, regular black holes are constructed by modifying the mass function in the metric to ensure that curvature invariants remain finite throughout the spacetime. Such constructions can be motivated by various theoretical approaches, including nonlinear electrodynamics~\cite{Ayon-Beato:2000mjt,Bronnikov:2022ofk,Capozziello:2024ucm}, loop quantum gravity~\cite{Modesto:2004xx}, and phenomenological geometric modifications~\cite{Fan:2016hvf,Maeda:2021jdc}. 

However, the theoretical realization of physically viable regular black holes is severely constrained by a series of fundamental no-go theorems. Foremost, following Penrose's singularity theorem, any regular black hole model inherently requires the violation of the strong energy condition within its core region~\cite{Penrose:1964wq,Maeda:2021jdc,Lan:2023cvz,Lan:2022bld}. Furthermore, within the framework of general relativity coupled to nonlinear electrodynamics, Bronnikov's theorem precludes the existence of static, spherically symmetric purely electric regular black holes, restricting viable solutions strictly to magnetic or dyonic configurations~\cite{Bronnikov:2000yz,Bronnikov:2017tnz,Cano:2020ezi,Bokulic:2022cyk,Bokulic:2023afx,DeFelice:2024seu,Bokulic:2025brf}. 
Furthermore, it has been demonstrated that regular black holes cannot be sourced by isotropic fluids if they are to maintain asymptotic flatness, necessitating anisotropic pressure profiles~\cite{Bronnikov:2017kvq}. Recent studies also suggest that standard spherical topologies are fundamentally prone to Cauchy horizon instabilities unless non-trivial topological structures are invoked~\cite{Calza:2025mrt}. 
Most critically, models featuring an inner Cauchy horizon are typically plagued by mass inflation, rendering their inner geometry dynamically unstable under small perturbations~\cite{Carballo-Rubio:2018pmi,Carballo-Rubio:2021bpr}. Confronted by these restrictive theorems, 
for further research on regular black holes, it is necessary to find systematic guiding principles and corresponding generalization methods without ad hoc assumptions.

Meanwhile, the detection of gravitational waves from binary black hole mergers~\cite{LIGOScientific:2016aoc} and the imaging of black hole shadows by the Event Horizon Telescope~\cite{EventHorizonTelescope:2019ggy,Vagnozzi:2022moj} have ushered in a new era of testing gravity in the strong-field regime. 
These breakthroughs enable observational access to near-horizon geometry and offer unprecedented opportunities to confront theoretical models with astrophysical data. 
In this context, regular black holes emerge as compelling alternatives to classical singular solutions and provide a natural platform for exploring potential quantum gravitational effects near the horizon.

In this work, we propose a \textit{finite curvature approach} to construct regular black hole spacetimes. 
Rather than beginning with a specific matter source \cite{Chamseddine:2016ktu} or a modified action \cite{Bueno:2024dgm}, we prescribe analytic forms for selected curvature invariants, particularly the Ricci and Weyl scalars, and solve the resulting differential equations for the metric function. 
The reason we use the Ricci and Weyl scalars instead of the traditional Kretschmann scalar is as follows:
1. In our model, the divergence and convergence properties of the Kretschmann scalar can be converted into those of the Ricci and Weyl scalars.
2. For the cases we consider, the differential equations derived from the Ricci and Weyl scalars are easier to solve, whereas the nonlinear equations derived from the Kretschmann scalar are difficult to solve analytically, which would prevent us from obtaining an analytic regular black hole metric.
This method guarantees regularity by construction and provides analytic control over the resulting geometric and physical properties.

We explored two main types of models: one based on a \textit{single bell-shaped curvature function} (such as Gaussian functions, hyperbolic secant functions, and rational functions), and the other based on the \textit{combination} of such functions.
These profiles are designed to satisfy essential physical criteria, including regularity at the origin, asymptotic flatness, and compatibility with standard energy conditions. 
Depending on the model parameters, the solutions can describe regular black holes with one or two horizons.

To assess the dynamical stability of these spacetimes, we investigate their \textit{quasinormal modes} (QNMs) under gravitational perturbations.
QNMs characterize the damped oscillatory response of a black hole, with frequencies determined by the background geometry \cite{Kokkotas:1999bd,Berti:2009kk,Konoplya:2011qq,Pedrotti:2024znu}. 
As the dominant feature of the ringdown phase in gravitational wave signals, QNMs serve as a powerful diagnostic of black hole structure and potential quantum gravity corrections \cite{Bhagwat:2017tkm,Abedi:2016hgu,LIGOScientific:2016aoc}.
Our analysis reveals that the \textit{shape of the effective potential}, specifically its width, asymmetry, and curvature, plays a critical role in shaping the QNM spectrum and late-time instabilities.

This paper is organized as follows. In Sec.~\ref{sec:finite_curvature}, we introduce the finite curvature framework based on Ricci and Weyl invariants. 
In Sec.~\ref{sec:models}, we construct explicit models of regular black holes using various curvature profiles. Sec.~\ref{sec:QNMs} presents a detailed analysis of the QNMs and their connection to the corresponding effective potentials. 
We conclude in Sec.~\ref{sec:conclusion} with a summary of our findings and a discussion of future directions.

\section{Finite curvature approach to regular black holes}
\label{sec:finite_curvature}

The complete set of curvature invariants for a Riemannian manifold consists of seventeen elements, known as Zakhary-McIntosh (ZM) invariants \cite{Zakhary:1997xas,Lan:2023cvz}. 
These invariants form a group of relations, termed syzygies\cite{Overduin:2020aiq}. 
One of such relations is the scalar form of the Ricci decomposition \cite{Weinberg:1972kfs}. 
In practice, for symmetric spacetimes, it is often sufficient to examine only the Kretschmann scalar, defined as $K = R_{\alpha\beta\mu\nu}R^{\alpha\beta\mu\nu}$, to determine whether a curvature singularity exists or not. 
Here, the Kretschmann scalar is more complex than other common curvature invariants for spherically symmetric spacetimes, such as the contractions of the Weyl tensor $W = W_{\alpha\beta\mu\nu}W^{\alpha\beta\mu\nu}$, the Ricci tensor $S = R_{\alpha\beta}R^{\alpha\beta}$, and the Ricci scalar $R$. 
This complexity arises from its construction, which includes more terms built from the metric components, some of which are nonlinear.

In this work, we focus on the case of black holes with the algebraic property $[(1,1)(11)]$\footnote{The algebraic property [(1,1)(11)] denotes the Segre type of the Ricci tensor, implying a diagonal form with grouped eigenvalues, typical for spherically symmetric metrics with $T^t_t \propto T^r_r$ \cite{Lan:2023cvz}.}, whose metric is given by
\begin{equation}
\label{eq:metric}
\dif s^2 = -f(r) \dif t^2 + f^{-1}(r) \dif r^2 + r^2 \dif \Omega^2,
\end{equation}
where $\dif \Omega^2 = \dif \theta^2 +\sin^2\theta\dif\phi^2$ and the metric function $f(r)$ is
\begin{equation}\label{Eq:metric-function}
f(r) = 1- \frac{2m(r)}{r}.
\end{equation}
Here,  $m(r)$  is a function of the radial coordinate $r$. 
To examine the curvatures, we take a syzygy from Ricci decomposition \cite{Weinberg:1972kfs}
\begin{equation}
\label{eq:ricci-decom}
W_{\alpha\beta\mu\nu}W^{\alpha\beta\mu\nu}+2R_{\alpha\beta}R^{\alpha\beta}
= R_{\alpha\beta\mu\nu}R^{\alpha\beta\mu\nu}
+\frac{1}{3}R^2.
\end{equation}
For the metric Eq.\ \eqref{eq:metric}, the corresponding curvature invariants are
\begin{subequations}
\label{eq:curvatures}
\begin{equation}
W=
\frac{4}{3 r^6} \left[r \left(r m''-4 m'\right)+6 m\right]^2,
\end{equation}
\begin{equation}
S = \frac{2 r^2 m''^2+8 m'^2}{r^4},
\end{equation}
\begin{equation}
R= \frac{2 r m''+4 m'}{r^2},
\end{equation}  
\begin{equation}
K=\frac{4 \left[r \left(r m''-2 m'\right)+2 m\right]^2}{r^6}+
\frac{16 \left(m-r m'\right)^2}{r^6}
+\frac{16 m^2}{r^6}
\end{equation}
\end{subequations}
where the prime denotes the derivative with respect to $r$.
From these expressions, we can make several conclusions:
\begin{enumerate}
	\item Only the Kretschmann scalar $K$ and the Weyl scalar $W$ contain $m(r)$ and its derivatives $m'(r)$, $m''(r)$.
    \item For metric Eq.\ \eqref{eq:metric}, $K$ is always non-negative (see also Eq.\ (10) in \cite{Bokulic:2025brf}); inequalities $R^2<4S<6K$ hold (sec.\ 6.1 in \cite{Dragasevic:2025mtf}), with $S$, $K$, $W$ expressible non-negatively
    \footnote{Furthermore, one can also prove that $K>W$ via this inequality and the Ricci decomposition. That is, for a black hole, if its $K$ is finite, then the other three scalars must be finite, whereas the converse statement does not hold.}.
    \item The Kretschmann scalar $K$, the Weyl tensor contraction $W$ and the Ricci tensor contraction $S$ include nonlinear terms involving  $m(r)$,  $m'(r)$, and $m''(r)$.
\end{enumerate}

The finite curvature approach to constructing regular black holes begins by specifying a form for the curvature and deriving the corresponding metric.
In particular, for Eq.\ \eqref{eq:curvatures}, once a form for the curvature invariants $W$, $K$, $R$, or $S$ is chosen, the associated second-order differential equation for the mass function $m(r)$ can be solved.
Although the Kretschmann scalar provides detailed insight into singularities, it is not the most convenient choice here. 
Its nonlinear structure leads to complicated second-order differential equations for $m(r)$, making the analysis unwieldy.

To simplify the calculations, it is important to select curvature invariants that are linear in $m(r)$ and its derivatives.
The Ricci scalar $R$ satisfies this condition, as it does not introduce nonlinear terms.
Similarly, while the Weyl tensor $W$ involves a squaring operation, the expression inside the square is itself linear, allowing it to be treated effectively within this framework.
In the remainder of this section, we will consider constructions based on each of these two curvature invariants.

\subsection{Starting from Ricci scalar curvature}
\label{sec:ricci}

Give a undetermined function $\beta(r)$, we can set 
\begin{equation}
\label{eq:master-R}
    R=2 \beta(r).
\end{equation} 
To keep the manifold regular, $\beta(r)$ must remain finite and be free of singularities across the domain $r\in(-\infty, \infty)$. 
Here we consider negative values of $r$ because singularities on the negative axis, $r\in(-\infty,0)$, or essential singularities at $r=0$, such as $\beta(r) = \me^{-r_0/r}$ for $r_0 > 0$, would both result in geodesic incompleteness\cite{Zhou:2022yio}, contradicting the definition of a regular manifold\cite{Hawking:1973uf}.

\begin{figure}[!ht]
\centering
\includegraphics[width=0.6\textwidth]{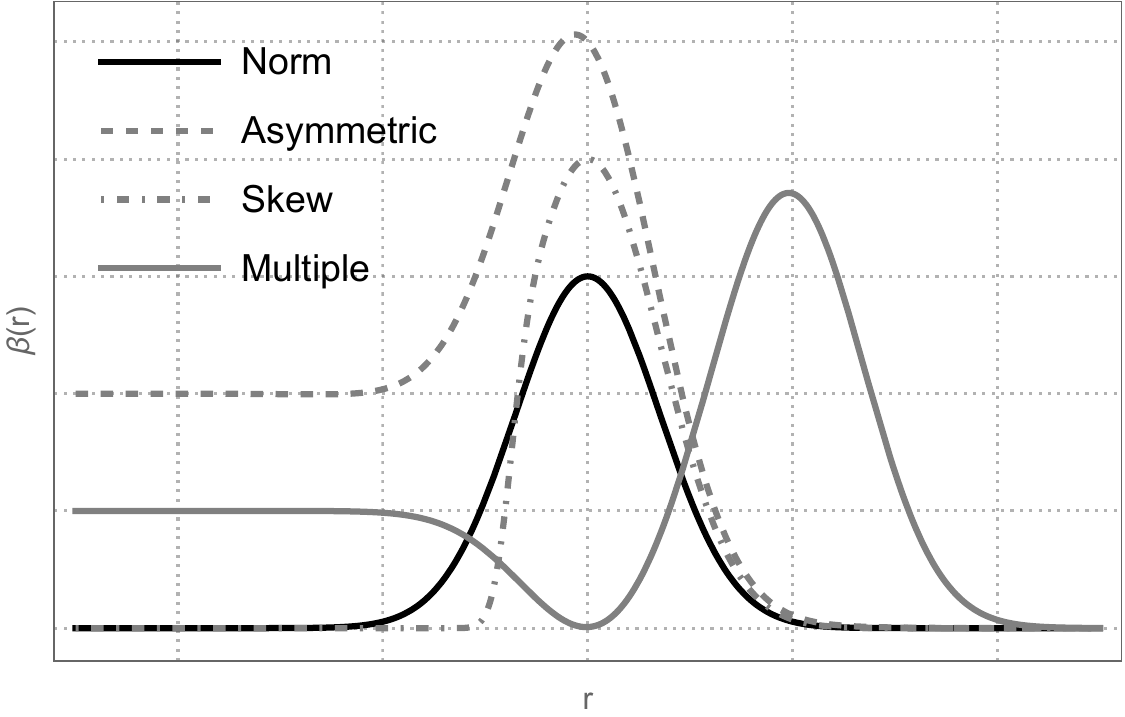}
\captionsetup{width=0.9\textwidth}
\caption{Schematic of the $\beta$ functions. The black line is a normal Gaussian function $C_1 \me^{-(x-x_0)^2}$, the gray dashed line is an asymmetric Gaussian function $C_1 \me^{-(x-x_0)^2}+C_2 [\tanh (-x)+1]$, the gray dot-dashed line is a Skew Gaussian function $C_1 \me^{-(x-x_0)^2}\int \dif x\;\me^{-(x-x_0)^2} $, and the gray solid line is a combination of multiple Gaussian functions $C_1\me^{-(x-x_0)^2} - C_2 \me^{-(x-x_1)^2}+C_3 [\tanh (x_1-x)+1]$.}
\label{fig:beta-shape}
\end{figure}

For the metric given in Eq.\ \eqref{eq:metric}, substituting Eq.\ \eqref{eq:master-R} yields
\begin{equation}
\frac{ r m''+2 m'}{r^2}=\beta(r),
\end{equation}
which has a general solution
\begin{equation}
\label{eq:sol-m-beta}
m(r)=c_1-\frac{c_2}{r}+
\int_0^r \frac{\dif y}{y^2} \int_0^y \dif x\; x^3 \beta(x), 
\end{equation}
where $c_1$ and $c_2$ are integration constants. 
For a regular black hole,  $m(r)$ must behave asymptotically as  $m(r) \sim O(r^n)$  with  $n \ge 3$  as  $r \to 0$ \cite{Bronnikov:2006fu,Fan:2016hvf}. 
Therefore, the regular condition requires that $c_1$  and  $c_2$  must vanish, i.e.,  $c_1 = c_2 = 0$, and $\beta(x) \sim O(r^{n})$ with $n\ge 0$ as $r\to 0$.
Meanwhile, the integral in $m(r)$ must finite as $r\to\infty$
\begin{equation}
\label{eq:anchor}
    \int_0^\infty \frac{\dif y}{y^2} \int_0^y \dif x\; x^3 \beta(x)<\infty, 
\end{equation}
such that regular black holes are asymptotically flat. 
This asymptotically flat condition requires that $\beta\sim O(r^{-n})$ with $n> 2$ as $r\to\infty$.
These conditions imply that  $\beta(r)$  can be realized as a combination of bell-shaped functions and sigmoid functions, see Fig.\ \ref{fig:beta-shape}.

Except for the Ricci scalar, additional curvature invariants in Eq.\ \eqref{eq:ricci-decom} are needed to verify regularity comprehensively.
Substituting Eq.\ \eqref{eq:sol-m-beta} with $c_1 = c_2 =0$ into the curvature invariants and analyzing their asymptotic behavior at $r=0$, we find
\begin{subequations}
\begin{equation}
\lim_{r\rightarrow{0}}K= \frac{2 \beta (0)^2}{3}+\frac{4}{3} \beta (0) r \beta '(0)+O\left(r^2\right),
\end{equation}
\begin{equation}
\lim_{r\rightarrow{0}}S= \beta (0)^2+2 \beta (0) r \beta '(0)+O\left(r^2\right),
\end{equation}
\begin{equation}
\lim_{r\rightarrow{0}}W= \frac{1}{75} r^2 \beta '(0)^2+O\left(r^3\right).
\end{equation}
\end{subequations}
Since $\beta(r)$ is nonsingular over the real axis, in particular at $r=0$, all four curvature invariants remain finite.

Next, we consider the geometric energy conditions \cite{Curiel:2014zba,Kontou:2020bta} derived from the Einstein tensor $G^{\mu}_{\;\;\nu}$ to guarantee that the constructed regular black holes are physical \cite{Maeda:2021jdc}. 
These energy conditions are defined as follows:
\begin{enumerate}
    \item Null energy condition (NEC): $-G^{0}_{\;\;0}$+$G^{i}_{\;\;i}\ge 0$, $i=1,2,3$;
    \item Weak energy condition (WEC): $-G^{0}_{\;\;0}\ge 0$ and $-G^{0}_{\;\;0}+G^{i}_{\;\;i}\ge 0$; 
    \item Strong energy condition (SEC):
$-G^{0}_{\;\;0}+\sum_{i}G^{i}_{\;\;i}\ge 0$
    and $-G^{0}_{\;\;0}+G^{i}_{\;\;i}\ge 0$;
    \item Dominant energy condition (DEC):
    $-G^{0}_{\;\;0}\ge 0$ and $-G^{0}_{\;\;0}\ge |G^{i}_{\;\;i}|$.
\end{enumerate}
where repeated indices do not imply summation.

The DEC cannot be violated, as it requires not only non–negative energy density as measured by any timelike observer, 
but also that the associated energy–momentum flux vector remain timelike or null, 
thereby ensuring that energy does not propagate faster than light.
It is a fundamental physical requirement for a regular black hole, see Ref.\ \cite{Maeda:2021jdc,Lan:2022bld}.
If this condition is substantially violated, the effective energy flux 
$J^\mu = -T^{\mu\nu}u_\nu$ may become spacelike for some observers, 
signalling superluminal energy transport and a potential breakdown of the standard causal structure.

Violation of the SEC has a distinct geometric and dynamical significance \cite{Lan:2023cvz}. 
The SEC requires that, for any timelike vector $u^\mu$, the Ricci contraction 
$R_{\mu\nu}u^\mu u^\nu \ge 0$, 
which, through the Einstein equations, implies that gravity is everywhere attractive 
for timelike observers. 
This condition plays a central role in the Hawking–Penrose singularity theorems, 
where it guarantees the focusing of timelike geodesic congruences via the Raychaudhuri equation.

If the SEC is violated, the term $R_{\mu\nu}u^\mu u^\nu$ may become negative, 
leading to a defocusing effect in the Raychaudhuri equation and thereby preventing 
the formation of caustics within a finite proper time. 
Geometrically, this corresponds to the emergence of an effective repulsive gravitational component, 
often associated with a de Sitter–like core in regular black hole models. 
Such repulsive behavior can halt geodesic convergence and remove the conditions required 
for singularity formation.

Unlike the dominant energy condition, however, violation of the SEC does not by itself 
imply superluminal energy fluxes or immediate causality violations. 
Indeed, many physically relevant systems—including scalar fields with positive potentials 
and effective descriptions of vacuum energy—naturally violate the SEC 
without exhibiting pathological behavior. 
In the context of regular black holes, controlled SEC violation in a compact central region 
is therefore not only expected but necessary in order to evade the classical singularity theorems, 
while maintaining consistency with asymptotic flatness and large-scale gravitational dynamics.

Now, substituting the metric, we obtain three inequalities for NEC, WEC, and DEC
\begin{subequations}
\label{eq:energy-cond}
\begin{equation}
    4 \int\dif r\; r^3 \beta (r) \ge r^4 \beta (r),
\end{equation}
\begin{equation}
    \int \dif r\; r^3 \beta (r) \ge 0,
\end{equation}
and 
\begin{equation}
    \left| 2 \int \dif r\;r^3 \beta (r) -r^4 \beta (r)\right| \le 2 \int \dif r\;r^3 \beta (r) ,
\end{equation}
\end{subequations}
while the SEC leads to
\begin{equation}
\label{eq:strong}
    2 \int r^3 \beta (r) \, dr\geq r^4 \beta (r).
\end{equation}
The inequalities in Eq.\ \eqref{eq:energy-cond} impose constraints on $\beta$, which must be satisfied for a physically regular black hole. 
According to Penrose’s singularity theorem, the last inequality Eq.\ \eqref{eq:strong} is expected to be violated in certain regions of spacetime. 
This violation suggests the presence of a repulsive interaction in the corresponding regions.

Here we propose a general method to construct regular black holes by requiring a finite Ricci scalar R. This construction is not restricted to Einstein gravity, but can also be applied to modified gravity theories, such as $f(R)$ gravity \cite{Hu:2023iuw} and other higher-derivative theories. Consequently, Eq.~\eqref{eq:master-R} does not provide direct information about the matter energy–momentum tensor. If one restricts attention to Einstein gravity only, the vacuum field equations $R_{\mu\nu}=0$, together with the uniqueness theorem, admit only the Schwarzschild solution, and Ricci flatness ($R=0$) cannot yield a regular black hole. In this sense, our regular black hole is not a vacuum solution of the classical Einstein equations, although it cannot be excluded that it may correspond to a vacuum solution in more general theories of gravity.

\subsection{Starting from Weyl scalar curvature}
\label{sec:weyl}

In this subsection, we construct regular black holes starting from a finite Weyl scalar curvature
\begin{equation}
\label{eq:master-W}
W = \frac{4}{3} \sigma^2(r),
\end{equation}
where the $\sigma(r)$ function is supposed to have the following properties:
\begin{enumerate}
\item The $\sigma(r) \to 0$ as $r\to \infty$, i.e., the Weyl scalar vanishes at infinity; 
\item  The $\sigma(r)$  no singularities in the region $r\in (-\infty, \infty)$.
\end{enumerate}
Substituting the metric Eq.\ \eqref{eq:metric} into Eq.\ \eqref{eq:master-W}, we arrive at
\begin{equation}
\label{eq:master-W-m}
\frac{m''}{r}-\frac{4 m'}{r^2}+\frac{6 m}{r^3}=\mp\sigma(r),
\end{equation}
whose general solution has the following form
\begin{equation}
\label{eq:sol-w}
m_\mp(r) =c_3 r^2+c_4 r^3\pm r^2 \int^r_0 \dif x \;\sigma(x) \mp r^3 \int^r_0 \frac{\dif x}{x} \sigma(x),
\end{equation}
where $c_3$ and $c_4$ are integration constants.
For the black hole to be regular at the origin, the mass function must behave as $m(r) \sim O(r^3)$ as $r \to 0$. This condition requires that the terms
\begin{equation}
\label{eq:anchor_2}
c_3 r^2, \quad r^3 \int \frac{\mathrm{d}r}{r} \sigma(r)
\end{equation}
vanish near $r=0$. Consequently, we must set $c_3 = 0$ (or arrange for it to be canceled by contributions from the integral terms), and impose $\sigma(0) = 0$. 
The latter condition implies that $\sigma(r)$ must vanish at least linearly as $r \to 0$, i.e., $\sigma(r) \sim r^n$ with $n \geq 1$.

We now consider the asymptotic behavior of the function $\sigma(r)$ as $r\to\infty$. According to Eq.\ \eqref{eq:sol-w}, the leading-order behavior of the mass function $m(r)$ at large $r$ must decay faster than or equal to a constant. In other words, $m(r)$ should scale with a negative power of $r$ as $r\to\infty$, such as $1/r$, $1/r^2$, and so on. 
This condition is essential to ensure that the spacetime remains asymptotically flat and that the time and radial coordinates maintain their correct physical roles, i.e., the metric retains the signature $(-+++)$, rather than flipping to $(+-++)$.

To analyze this, let us assume $\sigma(r) \sim r^k$ for large $r$ and substitute this into Eq.\ \eqref{eq:sol-w}. 
If $k \ge 0$, undesirable terms arise. For instance, when $k = 0$, the term $r^3 \int \sigma/r\dif r$ contributes a logarithmic divergence $\sim r^3 \ln r$, which cannot be canceled by the remaining terms. Similarly, if $k \ge 1$, the term $r^2 \int \sigma  \dif r$ introduces a leading-order contribution $\sim r^{k+2}$, which again dominates and disrupts asymptotic flatness.

\begin{figure}[!ht]
\centering
\includegraphics[width=0.6\textwidth]{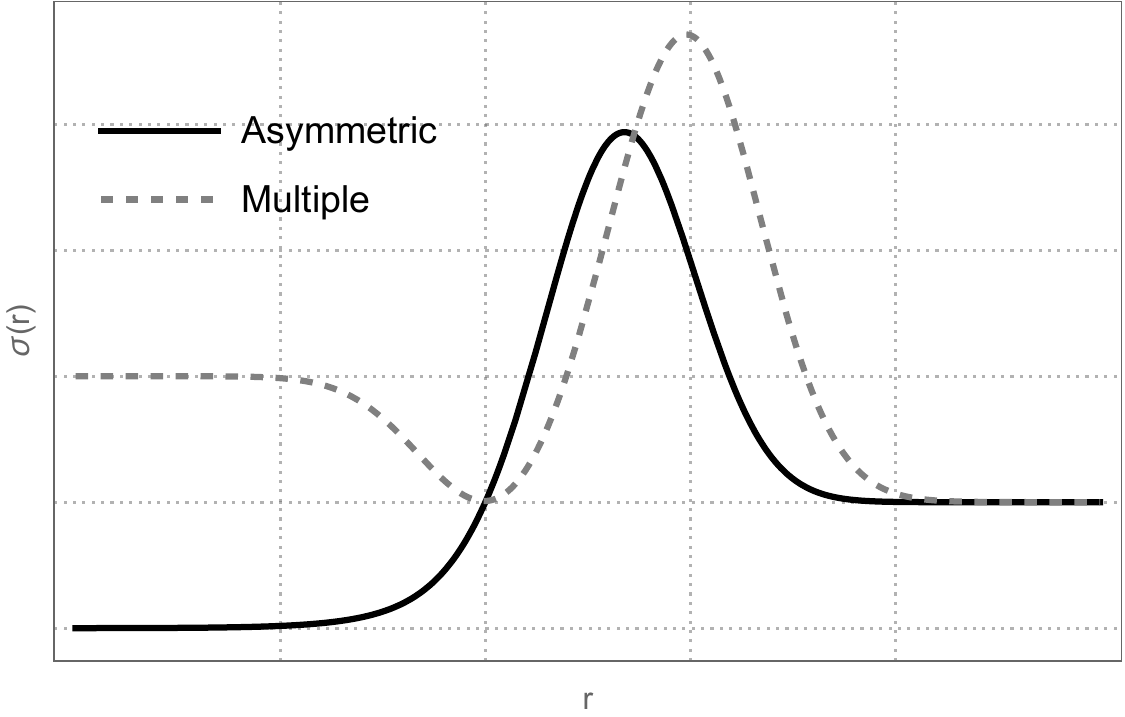}
\captionsetup{width=0.9\textwidth}
\caption{Schematic of the sigma functions. The black line is an asymmetric Gaussian function $C_1 \me^{-(x-x_0)^2}+C_2 [\tanh (-x)+1]$, and the gray dashed line is a combination of multiple Gaussian functions $C_1\me^{-(x-x_0)^2} - C_2 \me^{-(x-x_1)^2}+C_3 [\tanh (x_1-x)+1]$.}
\label{fig:beta-shape-2}
\end{figure}

Even for $k = -1$ or $k = -2$, the resulting behavior fails to satisfy the required conditions. We find that only when $k \le -3$, the leading-order terms decay sufficiently fast to maintain asymptotic flatness. Thus, the function $\sigma(r)$ must fall off at least as fast as $1/r^3$ for the spacetime to exhibit the correct asymptotic structure.
The Fig.\ \ref{fig:beta-shape-2} shows the possible forms of $\sigma(r)$ functions. 
The curvature invariants have the following asymptotic behaviors as $r\to 0$,
\begin{equation}
R\sim 24 c_4+O\left(r^1\right),\quad
K\sim 96 c_4^2+O\left(r^1\right),
\end{equation}
\begin{equation}
S\sim 144 c_4^2+O\left(r^1\right),\quad
W = \frac{4}{3} r^2 \sigma'(0)^2+O\left(r^3\right).
\end{equation}

Considering the energy conditions under this model, we obtain the components of the Einstein tensor
\begin{subequations}
\begin{equation}
G^{0}_{\;\;0}=G^{1}_{\;\;1}
=-\frac{4  }{r}\int\dif r\, \sigma (r)+6 \left(\int \dif r\frac{\sigma (r)}{r} - c_4\right),
\end{equation}
\begin{equation}
G^{2}_{\;\;2}=G^{3}_{\;\;3}
=\sigma (r)-\frac{2 }{r} \int\dif r\, \sigma (r) +6 \left(\int\dif r \frac{\sigma (r)}{r} -c_4\right).
\end{equation}
\end{subequations}
Here we restrict our discussion to the $m_-$ case, since the $m_+$ case follows directly by changing the sign of 
$\sigma$.
Thus, the null energy condition is given by 
\begin{equation}
-G^{0}_{\;\;0}+G^{2}_{\;\;2}\ge 0,
\end{equation}
which gives
\begin{equation}
r \sigma (r)+2 \int \sigma (r) \, \dif r\ge 0.
\end{equation}
The weak energy condition, $-G^0_{\;\;0}\ge 0$, gives additional inequality 
\begin{equation}
\int\dif r\; \frac{\sigma (r)}{r}  \le \frac{2}{3 r} \int \dif r \; \sigma (r) + c_4.
\end{equation}
On the basis of the above conditions, the strong energy condition provides an additional condition $-G^0_0+\sum_i G^i_i\ge0$, which is specifically expressed as
\begin{equation}
\label{eq:strong_weyl}
\frac{2 }{r} \int \dif r\;\sigma (r) +6 c_4\le \sigma (r)+6 \int\dif r\; \frac{\sigma (r)}{r}.
\end{equation}
Finally, the dominant energy condition is given by
\begin{equation}
\label{eq:dom_weyl1}
\left| 3 c_4-3 \int\dif r\; \frac{\sigma (r)}{r} +\frac{2 }{r}\int\dif r\; \sigma (r) \right| 
\le -3 \int\dif r\; \frac{\sigma (r)}{r}+\frac{2 }{r}\int \dif r\;\sigma (r) +3 c_4,
\end{equation}
\begin{equation}
\label{eq:dom_weyl2}
\left| -3 c_4+3 \int\dif r\; \frac{\sigma (r)}{r} +\frac{\sigma (r)}{2}-\frac{1}{r} \int\dif r\; \sigma (r) \right| 
\le -3 \int\dif r\; \frac{\sigma (r)}{r} +\frac{2 }{r} \int\dif r\; \sigma (r) +3 c_4.
\end{equation}
These inequalities, except the strong energy condition, are used to constrain the $\sigma(r)$, 
in particular the parameters in the $\sigma(r)$.

Summarily, while both Ricci- and Weyl-based constructions yield regular, asymptotically flat black holes, they differ in key aspects. 
\begin{enumerate}
    \item The Ricci approach (Sec.~\ref{sec:ricci}) starts from a linear differential equation $R=2\beta(r)$, leading to mass functions $m(r)$ via double integrals (Eq.~\eqref{eq:sol-m-beta}), with energy conditions (Eqs.~\eqref{eq:energy-cond} and \eqref{eq:strong} ) constraining $\beta(r)$ to bell-shaped forms for DEC compliance. 
    It naturally produces de Sitter cores but requires vanishing constants ($c_1=c_2=0$) for regularity.
    \item In contrast, the Weyl approach (Sec.~\ref{sec:weyl}) uses $W=4 \sigma^2(r)/3$, involving squared terms and yielding $m(r)$ with possible linear $\sigma(0)=0$ for origin regularity (Eq.~\eqref{eq:sol-w}). It allows non-zero $c_4$ for asymptotic adjustments but imposes faster decay ($\sigma\sim r^3$) and more complex DEC/SEC inequalities (Eqs.~\eqref{eq:strong_weyl}-\eqref{eq:dom_weyl2}). 
    \item Physically, Ricci models emphasize scalar curvature regularization, while Weyl models focus on tidal forces, leading to distinct effective potentials and QNMs (as explored in Sec.~\ref{sec:QNMs}). Both satisfy NEC/WEC in viable parameter ranges, highlighting complementary paths to singularity resolution.
\end{enumerate}

\subsection{Geodesic completeness and bell-shaped curvature profiles}

Geodesic completeness constitutes the definitive criterion for the absence of spacetime singularities.
While regular black holes are typically constructed so as to render curvature invariants finite, the boundedness of such scalars alone is insufficient to ensure geodesic completeness \cite{Zhou:2022yio,Lan:2023cvz}.
A spacetime can be regarded as genuinely nonsingular only if all inextendible geodesics admit an affine parametrisation over an infinite range.
In this subsection, we first establish the geodesic completeness of the regular black holes obtained within the framework developed in the preceding subsections.
We then discuss the physical motivations for adopting bell-shaped curvature profiles and summarize the theoretical constraints that such profiles must satisfy.

\vspace{0.3em}
\noindent\textit{Geodesic completeness.}—
We begin by analytically extending the radial coordinate of the metric in Eq.~\eqref{eq:metric} to the full real line, $r\in(-\infty,\infty)$.
The metric function $f(r)$ is constructed from rapidly decaying and analytic curvature profiles, which guarantees regularity throughout the manifold.
The resulting geometry describes a wormhole-like spacetime connecting two asymptotically flat regions through a smooth throat located at $r=0$.
Near the core, regularity of curvature invariants requires the mass function to behave as
\begin{equation}
    m(r) \sim \frac{1}{12}\,\beta(0)\, r^3 + \mathcal{O}(r^4),
    \qquad r \to 0 ,
\end{equation}
which in turn yields the expansion
\begin{equation}
    f(r) \approx 1 - \frac{\beta(0)}{6}\, r^2 > 0 .
\end{equation}
This behavior follows directly from the requirement of finite curvature invariants in the Ricci-scalar construction; see the discussion around Eq.~\eqref{eq:anchor}.
The Weyl-scalar construction leads to the same conclusion, since the leading $r^3$ behavior of $m(r)$ near the origin constitutes a special case of the Ricci-based approach; cf.~Eq.~\eqref{eq:anchor_2}.
The resulting effective geometry near $r=0$ is therefore de Sitter–like, allowing geodesics to pass smoothly through the throat.

Since the metric function reduces at $r=0$ to the same local form considered in Ref.~\cite{Hu:2023iuw}, the proof of geodesic completeness in the present setup closely parallels that given in Sec.~2.2 of that work.
In particular, by transforming to Painlev\'e–Gullstrand coordinates, one can explicitly verify that both null and timelike geodesics extend to infinite affine or proper time.
Furthermore, because the curvature profiles $\beta(r)$ and $\sigma(r)$ are smooth and free of singularities over the entire real line, the resulting spacetime differs qualitatively from the Hayward geometry, where geodesic incompleteness arises despite finite curvature invariants \cite{Zhou:2022yio}.
We therefore conclude that the regular black holes constructed via either of our two proposed approaches are necessarily geodesically complete.

\vspace{0.3em}
\noindent\textit{Bell-shaped curvature profiles and physical motivation.}—
As demonstrated in the preceding two subsections, the requirement of curvature regularity in nonsingular black hole geometries generically implies that curvature invariants exhibit a bell-shaped radial profile, at least in simple and symmetric settings.

In the context of regular black holes, bell-shaped curvature profiles—such as Gaussian, hyperbolic secant, or rational forms—are widely employed as phenomenological encodings of quantum-gravitational effects.
Their primary role is to ensure that curvature invariants, including the Ricci scalar $R$, the Weyl scalar $W$, and the Kretschmann scalar $K$, remain finite throughout spacetime, thereby resolving the curvature singularities present in classical general relativity solutions such as Schwarzschild or Kerr.
The underlying motivation arises from quantum gravity frameworks, in which classical point-like singularities are effectively smeared due to the presence of a fundamental minimal length or spacetime discreteness, while effective field theory and ultraviolet consistency impose additional constraints.

From the perspective of quantum gravity, it is widely expected that Planck-scale physics ($\ell_{\rm Pl}\simeq1.6\times10^{-35}\,\mathrm{m}$) modifies the short-distance structure of spacetime so as to prevent divergent curvatures.
Bell-shaped functions naturally capture this behavior by localizing quantum corrections near the central region ($r\simeq0$), while decaying sufficiently rapidly at large radii to preserve asymptotic flatness.
In loop quantum gravity, classical singularities are replaced by quantum bounces induced by nonperturbative effects \cite{Ashtekar:2023cod}, leading to effective black hole geometries in which curvature invariants display bell-like behavior, as observed for instance in self-dual LQG black hole solutions \cite{Modesto:2009ve}.
Gaussian-type profiles may be interpreted as a consequence of area quantization, which distributes the classical singularity over a finite Planck-scale region.

Phenomenological analyses further indicate that rotating LQG-inspired black holes exhibit photon spheres and shadows that are slightly smaller than those of Kerr black holes, with deviations controlled by the minimal area parameter $A_{\rm min}\simeq4.4\,\ell_{\rm Pl}^2$, see e.g.~Refs.~\cite{Perez:2017cmj,Bojowald:2020dkb}.
A closely related motivation arises in noncommutative geometry, where spacetime coordinates become noncommuting operators and point-like matter sources are replaced by smooth distributions.
In this framework, Gaussian matter profiles $\rho(r)\propto\exp(-r^2/\theta)$ yield regular black hole solutions with finite core densities \cite{Nicolini:2005vd,Alkac:2025bbb}, typically featuring de Sitter--like interiors consistent with expectations from quantum vacuum fluctuations \cite{Filho:2024zxx}.

Bell-shaped curvature profiles also arise in higher-derivative and nonlocal theories of gravity.
In ultraviolet self-complete or nonlocal gravity models, singularities are avoided through nonlocal interactions that effectively bound curvature invariants \cite{Modesto:2010uh,Spallucci:2011rn}.
Hyperbolic secant profiles often mimic quantum-corrected de Sitter cores. Rational profiles, on the other hand, resemble the phenomenological metric originally proposed by Bardeen \cite{Bardeen:1968nsg}, which was later demonstrated to emerge as an exact solution when gravity is coupled to nonlinear electrodynamics \cite{Ayon-Beato:1998hmi}.

From an effective field theory perspective, gravity is treated as a low-energy expansion supplemented by higher-dimensional operators.
For regular black hole solutions to remain under perturbative control, curvature profiles must not trigger a breakdown of the EFT expansion below the cutoff scale.
Causality constraints and the absence of superluminal propagation place bounds on admissible profiles \cite{deRham:2021bll}.
Phenomenological constraints from gravitational-wave time delays further require the characteristic smearing scale $\ell$ to satisfy $\ell/M\lesssim\mathcal{O}(0.1)$ for stellar-mass black holes \cite{Cardoso:2018ptl}.
Excessively broad profiles may induce ghost-like excitations or dynamical instabilities, rendering them theoretically unacceptable \cite{Deffayet:2021nnt}.

Unitarity additionally demands the absence of tachyonic or negative-norm modes in scattering amplitudes.
In the case of Gaussian profiles, this requirement typically enforces $\ell\ll M$, ensuring that the Schwarzschild limit is recovered in the weak-field regime.
Energy condition considerations also play an important role: although regular black holes necessarily violate the strong energy condition near their cores in order to generate effective repulsion, consistency requires that the weak and null energy conditions be restored asymptotically \cite{Visser:1999de}.
Appropriate choices of curvature profiles can minimize such violations, with rational profiles often exhibiting improved compliance with the dominant energy condition \cite{Balart:2014cga}.

Finally, in genuinely ultraviolet-complete theories of gravity, bell-shaped curvature profiles frequently emerge as effective descriptions of nonsingular black hole solutions.
In quadratic (Stelle) gravity, regular solutions arise with bounded curvature potentials \cite{Holdom:2022npq}, while gravitational-wave observations impose stringent constraints on the propagation speed of additional modes \cite{Frolov:2024hhe}.
Asymptotically safe gravity predicts fixed-point behavior of running couplings, from which effective bell-shaped profiles naturally follow, and nonlocal extensions smear classical singularities while preserving unitarity by avoiding additional propagator poles \cite{Bonanno:2025dry}.

Ho\v{r}ava gravity likewise admits regular black hole solutions with Gaussian-like curvature corrections, whose parameters are constrained by observations such as black hole shadow measurements of M$87^*$ \cite{Kehagias:2009is}.
Collectively, these frameworks resolve classical singularities while maintaining asymptotic flatness.
Observable consequences include modest shifts in quasinormal mode spectra, small reductions in black hole shadow sizes, and mild lensing time delays, all of which remain compatible with current bounds.
Although no-go theorems suggest that violations of energy conditions are unavoidable, semiclassical and quantum interpretations provide a natural resolution, supporting the view that regular black holes represent consistent quantum extensions of classical general relativity rather than indications of radical new physics \cite{Roupas:2022gee,Malafarina:2017csn}.

\section{Constructions of the regular black holes}
\label{sec:models}

Now, let us turn to the specific examples. 
We separate the discussions here into two cases according to the above general theory.
The first is dedicated to the Ricci-scalar approach, and the second is to the Weyl-scalar approach.

\subsection{Single bell-shaped functions}
\label{sec:single}

\subsubsection*{Gaussian functions}

The Gaussian functions are typical bell-shaped functions.
We suppose that the beta function in Eq.\ \eqref{eq:master-R} has the following form
\begin{equation}
\beta_1(r)=A \exp\left[-\frac{(r-r_0)^2}{s^2}\right],
\end{equation}
where the parameters satisfy
\begin{equation}
A>0,\quad
s>0,\quad
r_0\in\mathbb{R}.
\end{equation}
For simplicity, we set $r_0=0$, and obtain the mass term from the differential equation Eq.\ \eqref{eq:sol-m-beta} with $c_1=c_2=0$
\begin{equation}\label{eq:m1}
m_1(r)=\frac{A s^3}{4 r} 
 \left[\sqrt{\pp } r \,\text{erf}\left(\frac{r}{s}\right)+2 s \left(\me^{-\frac{r^2}{s^2}}-1\right)\right],
\end{equation}
where $\text{erf}(\cdot)$ is error function. 

The regular black hole obtained by our method may be interpreted either as an effective quantum correction to the classical vacuum solution or as a spacetime generated by some form of matter or field. Among various possibilities, the most straightforward interpretation is to regard it as sourced by an anisotropic fluid. Other common interpretations, such as those based on nonlinear electrodynamics, suffer from certain drawbacks: not all nonlinear electrodynamics models admit regular black hole solutions, and the corresponding actions that do so are typically not derived from first principles. Moreover, different matter interpretations generally lead to different perturbation equations in perturbative analyses, a point that will be discussed later in connection with quasinormal modes.

\begin{figure}[!ht]
\centering
\includegraphics[width=0.5\textwidth]{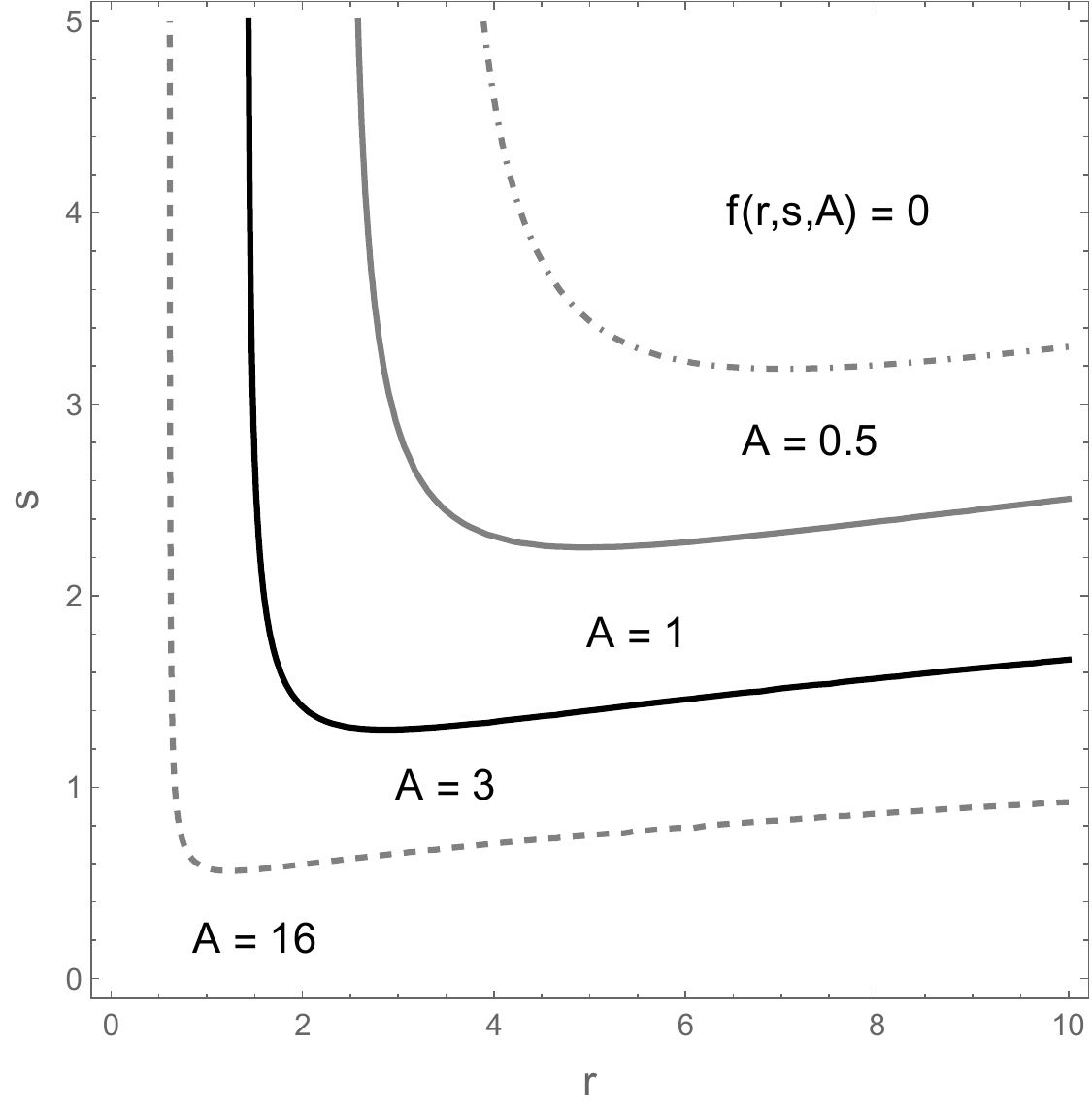}
\captionsetup{width=0.9\textwidth}
\caption{Horizons with different parameters: A model based on the Gaussian function. The curves in the figure are obtained based on Eqs.~\eqref{eq:m1} and ~\eqref{eq:horizon-deter}.}
\label{fig:horizon-gauss}
\end{figure}

The determining condition for the event horizon $r_{\rm H}$ of a black hole is that the metric function Eq.~\eqref{Eq:metric-function} equals zero,
\begin{equation}\label{eq:horizon-deter}
    f(r_{\rm H})=1-\frac{2m(r_{\rm H})}{r_{\rm H}}=0.
\end{equation}
Horizons with different parameters are shown in Fig.\ \ref{fig:horizon-gauss}.
It notes that this mode can have one horizon or two horizons according to the parameters. 

The asymptotic behavior of $m_1(r)$ at $r=0$ reads
\begin{equation}
m_1(r) \sim \frac{A}{12}r^3 + O(r^4),
\end{equation}
which meets the condition of a black hole being regular.
Meanwhile, the mass function $m_1(r)$ tends to a constant
\begin{equation}
\lim_{r\to\infty} m_1(r)  = \frac{\sqrt{\pp } }{4} A s^3,
\end{equation}
thus, this model is asymptotically flat.

Further, we use the energy conditions to fix the allowed domain parameters.
The null and weak energy conditions are given by
\begin{equation}
\label{eq:null}
\me^{-\frac{r^2}{s^2}} \left(r^4+2 r^2 s^2+2 s^4\right)-2 s^4\le 0.
\end{equation}
This inequality holds for all $r\ge 0$ and $s>0$. To prove this statement, we
denote the left hand of the inequality by $\epsilon_1(r)=\me^{-\frac{r^2}{s^2}} \left(r^4+2 r^2 s^2+2 s^4\right)-2 s^4$,
then we have
\begin{equation}
\epsilon_1(0) = 0,\quad
\lim_{r\to\infty} \epsilon_1(r) =-2 s^4<0,
\end{equation}
and
\begin{equation}
\frac{\dif \epsilon_1(r) }{\dif r}=-\frac{2 r^5 \me^{-\frac{r^2}{s^2}}}{s^2}<0.
\end{equation}
In other words, $\epsilon_1(r)$ is monotonicly decreasing function and $\epsilon_1(0)=0$. 
Therefore, $\epsilon_1(r)\le 0$ and the statement is proved.

Now let us consider the strong energy conditions, we have
\begin{equation}
\me^{-\frac{r^2}{s^2}} \left(r^4+r^2 s^2+s^4\right)-s^4 \le 0.
\end{equation}
To find the violation region, we denote $\epsilon_2(r)=\me^{-\frac{r^2}{s^2}} \left(r^4+r^2 s^2+s^4\right)-s^4$.
The $\epsilon_2(r)$ function at $r=0$ and infinity are
\begin{equation}
\epsilon_2(0)=0,\quad
\lim_{r\to\infty} \epsilon_2(r) = -s^4<0.
\end{equation}
The first derivative is
\begin{equation}
\epsilon'_2(r)=\frac{2 r^3 \me^{-\frac{r^2}{s^2}} \left(s^2-r^2\right)}{s^2},
\end{equation}
which has two different signs
\begin{equation}
\epsilon'_2(r)
\begin{cases}
>0, & r<s,\\
<0, & r>s.
\end{cases}
\end{equation}
This implies that $\epsilon'_2(r)$ is monotonically increasing function in the range $r<s$, 
while it monotonically decreases when $r>s$.
Furthermore, from $\epsilon'_2(s)=0$ and $\epsilon''_2(s)=-4s^2/\me<0$, we can deduce that $r=s$ is the maximum point of $\epsilon_2(r)$.
In addition,  there is a critical point $r_c$, such that $\epsilon_2(r)>0$ in the range $r<r_c$ and $\epsilon_2(r)<0$ in the range $r>r_c$. This critical point $r_c$ can be estimated from $\epsilon_2(r_c) =0$, which gives $r_c\approx 1.33913 s$.
Summarily, the strong energy condition is violated in the range
\begin{equation}
r<r_c,\quad \epsilon_2>0,
\end{equation}
and holds in the range
\begin{equation}\label{eq:rge0}
r\ge r_c,\quad \epsilon_2\le 0.
\end{equation}

At last, the dominant energy condition leads to 
\begin{equation}\label{eq:dominant-Gauss}
s^4\geq \left| s^4-\me^{-\frac{r^2}{s^2}} \left(r^4+s^2 r^2+s^4\right)\right| +s^2 \me^{-\frac{r^2}{s^2}} \left(r^2+s^2\right).
\end{equation}
We note that the formula is exactly $-\epsilon_2(r)$ in absolute operator brackets. 
In the range $r\ge r_c$, according to Eq.~\eqref{eq:rge0}, we rewrite Eq.~\eqref{eq:dominant-Gauss} as
\begin{equation}
r^4 \me^{-\frac{r^2}{s^2}}\ge 0,
\end{equation}
which holds for all $r\ge 0$ and $s>0$.
For the case $r< r_c$, Eq.~\eqref{eq:dominant-Gauss} reduces to Eq.\ \eqref{eq:null}, which is proved to be maintained.

In conclusion, when we adopt the Gaussian functions as beta functions to construct regular black holes, the null, weak, and dominant energy conditions of the black holes can be satisfied.
\subsubsection*{Hyperbolic secant functions}

\begin{figure}[!ht]
\centering
\includegraphics[width=0.5\textwidth]{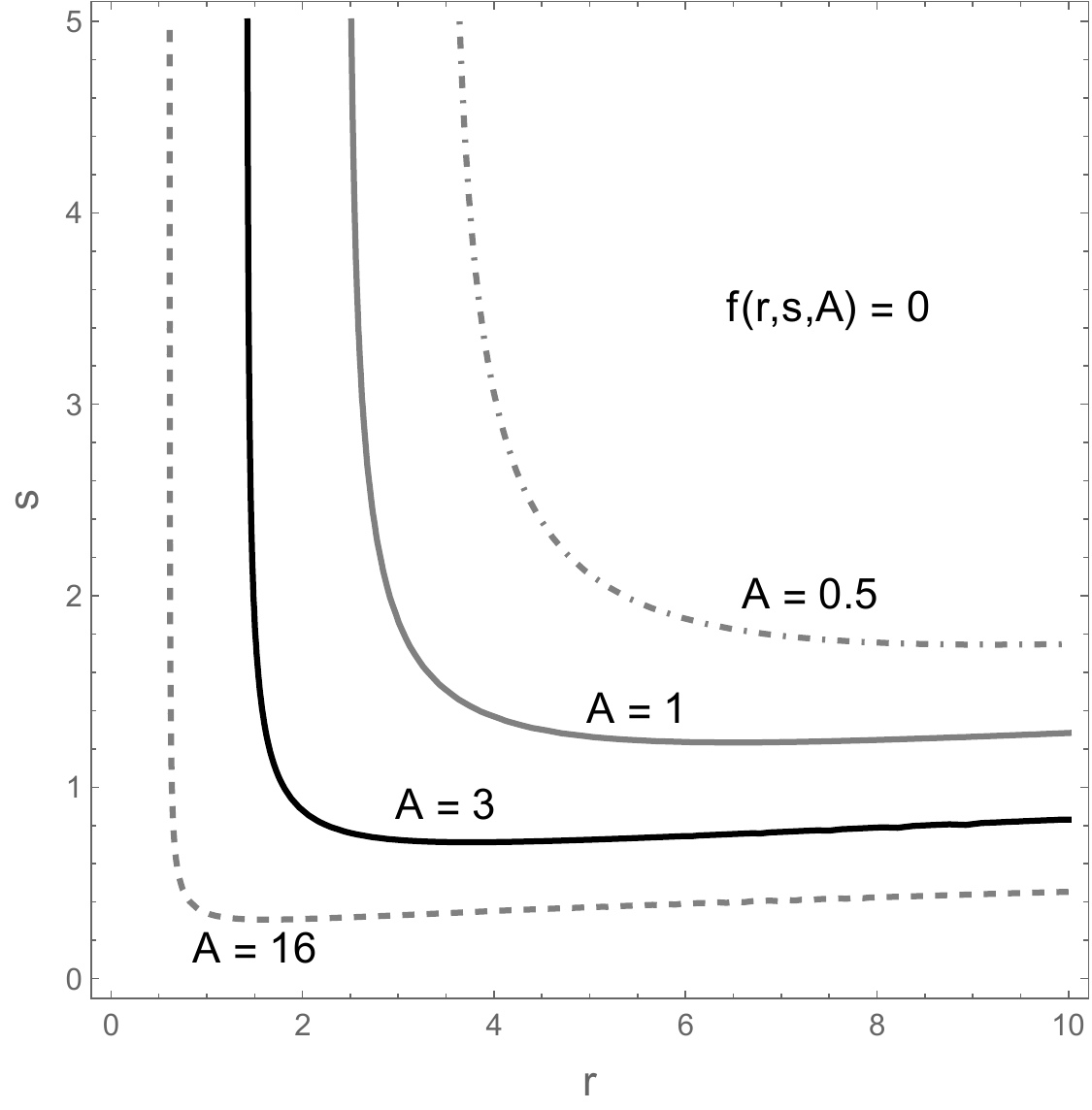}
\captionsetup{width=0.9\textwidth}
\caption{Horizons with different parameters: A model based on the hyperbolic secant function. The curves in the figure are obtained based on Eqs.~\eqref{eq:m2} and ~\eqref{eq:horizon-deter}.}
\label{fig:horizon-sech}
\end{figure}

Next, we consider the hyperbolic secant functions
\begin{equation}
\beta_2(r) = A \ \text{sech}^n\left[\frac{r-r_0}{s}\right].
\end{equation}
In the following, we concentrate on the situation with $n=1$ and  $r_0=0$.

\begin{equation}\label{eq:m2}
\begin{split}
m_2(r)  = &
\frac{3 A s^4 \me^{r/s} }{4 r}\Phi \left(-\me^{\frac{2 r}{s}},4,\frac{1}{2}\right)
-A s^3 \me^{r/s} \Phi \left(-\me^{\frac{2 r}{s}},3,\frac{1}{2}\right)\\
&+\frac{A r s^2 \me^{r/s} }{2} \Phi \left(-\me^{\frac{2 r}{s}},2,\frac{1}{2}\right)
+\frac{A s^4 }{128 r} \left[\psi ^{(3)}\left(\frac{3}{4}\right)-\psi ^{(3)}\left(\frac{1}{4}\right)\right]-\frac{\pp^3 A s^3}{8}. 
\end{split}
\end{equation}
where $\Phi(z,s,a)$ is the Lerch transcendent and can be represented via
polylogarithms $\text{Li}_n(z)$
\begin{equation}
2^{1-n} \me^{r/s} \Phi \left(-\me^{\frac{2 r}{s}},n,\frac{1}{2}\right)
=
\mi \left[\text{Li}_n\left(-\mi \me^{r/s}\right)-\text{Li}_n\left(\mi \me^{r/s}\right)\right],
\end{equation}
and $\psi^{(m)}(z)=\dif^{m+1}\ln \Gamma(z)/\dif z^{m+1}$ is polygamma function of order $m$, where $\Gamma(z)$ is the gamma function. The horizons with different parameters are shown in Fig.\ \ref{fig:horizon-sech}.

The asymptotic behaviors of the first three terms of Eq.~\eqref{eq:m2} as $r\to \infty$ are respectively
\begin{subequations}
\begin{equation}
\frac{A r s^2\me^{r/s}}{2}  \Phi \left(-\me^{\frac{2 r}{s}},2,\frac{1}{2}\right)
\sim A \pp s r^2,
\end{equation}
\begin{equation}
 A s^3 \me^{r/s} \Phi \left(-\me^{\frac{2 r}{s}},3,\frac{1}{2}\right)\sim 
2 A \pp s r^2+\frac{\pp ^3 s^3}{2},
\end{equation}
\begin{equation}
\frac{3 A s^4 \me^{r/s} }{4 r} \Phi \left(-e^{\frac{2 r}{s}},4,\frac{1}{2}\right)\sim 
A \pp s r^2+\frac{3 A \pp ^3 s^3}{4 }.
\end{equation}
\end{subequations}
Therefore, the limit of the first three terms of Eq.~\eqref{eq:m2} as they approach infinity is finite
\begin{equation}
\begin{split}
\lim_{r\to\infty} \Big[
\frac{3 A s^4 \me^{r/s} }{4 r}\Phi \left(-\me^{\frac{2 r}{s}},4,\frac{1}{2}\right)
&-A s^3 \me^{r/s} \Phi \left(-\me^{\frac{2 r}{s}},3,\frac{1}{2}\right)\\
&+\frac{A r s^2 \me^{r/s} }{2} \Phi \left(-\me^{\frac{2 r}{s}},2,\frac{1}{2}\right) 
\Big]
=\frac{\pp^3 A s^3}{4}.
\end{split}
\end{equation}
Finally, we find that $m_2(r)$ approaches a finite values as $r\to\infty$
\begin{equation}
\lim_{r\to\infty} m_2(r) =\frac{\pp ^3  A s^3}{4}+0-\frac{\pp ^3 A s^3}{8}=\frac{\pp ^3  A s^3}{8}.
\end{equation}
On the other side, as $r\to0$ we have
\begin{equation}
m_2(r)\sim \frac{A r^3}{12}+O\left(r^4\right),
\end{equation}
which also meets the condition of a black hole being regular. The energy conditions are shown in the following Fig.\ \ref{fig:ecd-sech},
\begin{figure}[!ht]
\centering
\includegraphics[width=0.5\textwidth]{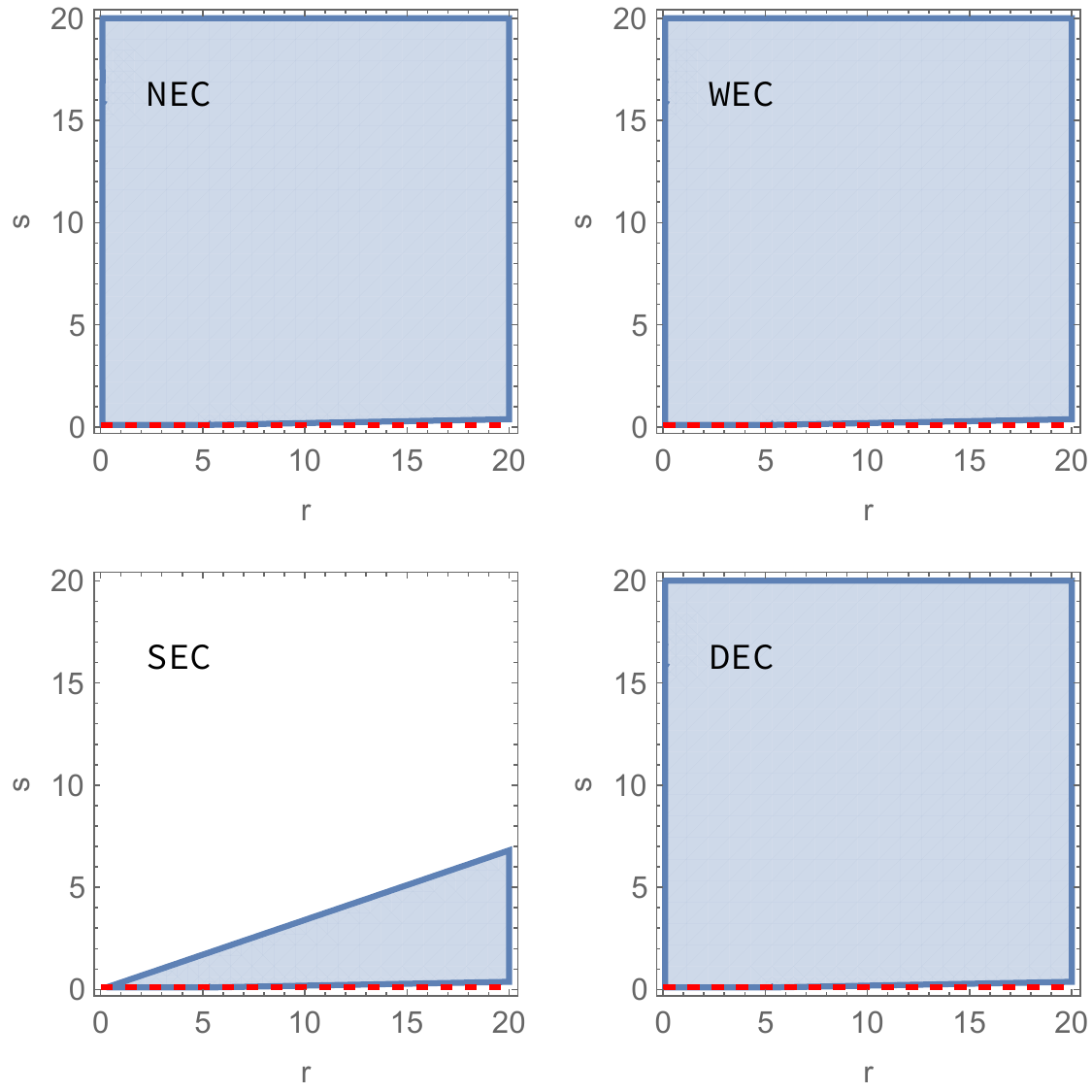}
\captionsetup{width=0.9\textwidth}
\caption{The energy conditions of regular black holes constructed by taking the hyperbolic secant functions as the beta functions. Among them, the shade areas display the parameter spaces that satisfy the energy conditions.}
\label{fig:ecd-sech}
\end{figure} 
where the shade areas display the parameter spaces that satisfy the energy conditions. 
It is noted from the Fig.\ \ref{fig:ecd-sech} that, first, there is a slight violation of all energy conditions in the lower-right range;
second, the SEC is violated in the upper-left range, which is consistent with our expectation; third, the DEC is satisfied across almost the entire range, meaning that the regular black hole we constructed is physically plausible.

\subsubsection*{Fuzzy logic functions}

The fuzzy logic functions also have a bell-shaped form. The general form is
\begin{equation}
\beta_3(r) = \frac{A}{1+(r-r_0)^{2n}/s^2}.
\end{equation}
Taking $n=1$ and $r_0=0$, we find that 
the integral in Eq.\ \eqref{eq:sol-m-beta} is divergent as $r$ when $r$ approaches to the infinity
\begin{equation}
m_3(r) \sim \frac{A r s^2}{2} -\frac{\pp A s^3}{2} +O\left(r^{-1}\right), \text{ as }  r\to \infty,
\end{equation}
Although it can lead to an asymptotically flat spacetime, it affects the strong energy condition.
Substituting the $\beta_3(r)$ with $n=1$ and $r_0=0$ into inequality of SEC, we obatin
\begin{equation}
r^2\ge \left(r^2+s^2\right) \ln \left(\frac{r^2}{s^2}+1\right),
\end{equation}
which is satisfied only for $r=0$. The proof method is similar to the one in the Gaussian function, i.e., 
we first define a function $\delta_0(r)=r^2-\left(r^2+s^2\right) \ln \left(r^2/s^2+1\right)$.
Then from $\delta_0(0)=0$ and $\delta_0'(r)=-2 r \ln \left(r^2/s^2+1\right)<0$, we can deduce that $\delta_0(r)$ is a monotonically decreasing function and $\delta_0(r)\le 0$. 
In other words, the SEC is completely violated except at the center $r=0$.

\begin{figure}[!ht]
\centering
\includegraphics[width=0.5\textwidth]{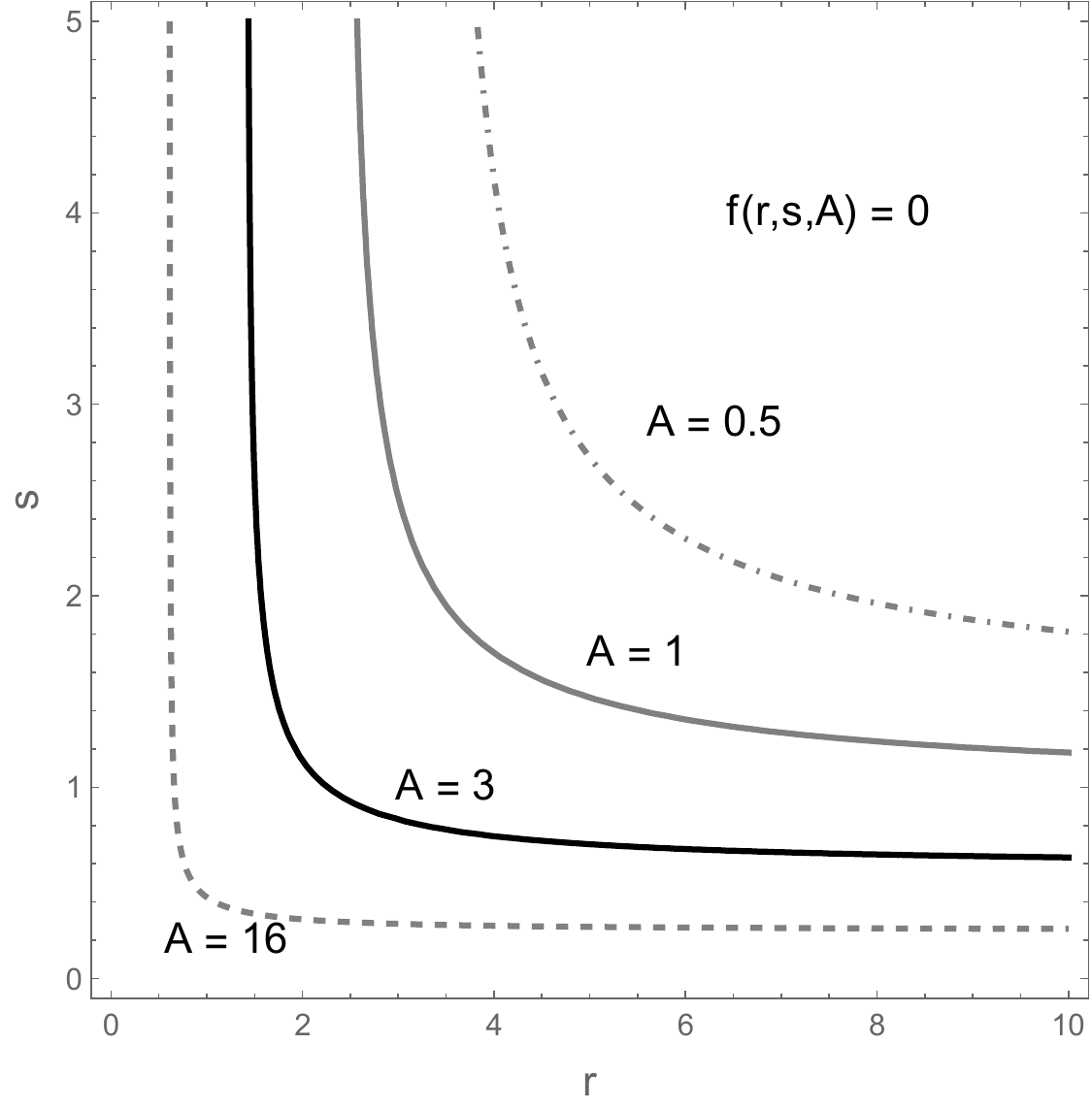}
\captionsetup{width=0.9\textwidth}
\caption{Horizons with different parameters: A model based on the fuzzy logic function. The curves in the figure are obtained based on Eqs.~\eqref{eq:m3} and ~\eqref{eq:horizon-deter}.}
\label{fig:horizon-fuzzy}
\end{figure}

Therefore, we consider the fuzzy logic function with $n=2$ in the following. 
The mass function for this case has an analytical expression
\begin{equation}
\begin{split}\label{eq:m3}
m_3(r)  =& -\frac{A s^4 }{4 r}\ln \left(\frac{r^4}{s^4}+1\right)
+\frac{A s^3 }{4 \sqrt{2}}\ln \left(\frac{r^2-\sqrt{2} r s+s^2}{r^2+\sqrt{2} r s+s^2}\right)\\
&+\frac{A s^3}{2 \sqrt{2}}
 \left[\arctan\left(\frac{\sqrt{2} r}{s}+1\right)+\arctan\left(\frac{\sqrt{2} r}{s}-1\right)\right].
\end{split}
\end{equation}
It approaches a constant as $r\to\infty$
\begin{equation}
\lim_{r\to\infty} m_3(r) =\frac{\pp  A s^3}{2 \sqrt{2}},
\end{equation}
and meet the condition of regular black holes as $r\to 0$, i.e.,
\begin{equation}
m_3(r)\sim \frac{A r^3}{12}+O\left(r^4\right).
\end{equation}
The horizons with different parameters are shown in Fig.\ \ref{fig:horizon-fuzzy}.
It notes that this model has only one horizon because $f(r)=1-2m_3(r)/r$ is a monotonically decreasing function.
To prove it, we will show that $f'(r)<0$ for $r>0$. The inequality $f'(r)<0$ gives
\begin{equation}
0<\xi(x) \equiv x\arctan\left(x\right)-\ln \left(x^2+1\right),\quad
x=r/s,
\end{equation}
thus if $\xi(0)=0$ and $\xi'(x)>0$ for $x>0$, $0<\xi(x)$ will be valid.
$\xi'(x)$ gives
\begin{equation}
\xi'(x)=\arctan(x) -\frac{x}{x^2+1},
\end{equation}
while $\xi'(x)$ is a monotonically increasing function, because $\xi''(x)=2x^2/(1+x^2)^2>0$ for $x>0$, 
and since $\xi'(0)=0$, we can deduce that $\xi'(x)>0$ for $x>0$. 
Summarily, we have $0<\xi(x)$, i.e., this model has only one horizon.

Now let us consider the NEC and WEC
\begin{equation}
r^4\le \left(r^4+s^4\right) \ln \left(\frac{r^4}{s^4}+1\right).
\end{equation}
This inequality holds in all parameter spaces. To prove it, 
we first construct a function
\begin{equation}
\delta_1(r)=\left(r^4+s^4\right) \ln \left(\frac{r^4}{s^4}+1\right)-r^4,
\end{equation}
and obtain
\begin{equation}
\delta_1(0)=0,\quad
\delta'_1(r) = 4 r^3 \ln \left(\frac{r^4}{s^4}+1\right)>0,
\end{equation}
which indicates that $\delta_1(r)$ is monotonically increasing function and $\delta_1(r)\ge 0$.

The SEC provides an inequality
\begin{equation}
2 r^4\le \left(r^4+s^4\right) \ln \left(\frac{r^4}{s^4}+1\right).
\end{equation}
After defining a function
\begin{equation}
\delta_2(r)=\left(r^4+s^4\right) \ln \left(\frac{r^4}{s^4}+1\right)-2 r^4,
\end{equation}
we can find a critical point $r_c$ which is
\begin{equation}
\delta_2(r_c) =0,\quad 
r_c \approx 1.40723 s.
\end{equation}
And
\begin{equation}
    \delta_2(r)
\begin{cases}
<0, & r<r_c,\\
>0, & r>r_c.
\end{cases}
\end{equation}
Therefore, the SEC is violated in the inner domain of the black hole $r<r_c$ as we expected.

Finally, the DEC results in  
\begin{equation}
\left| \ln \left(\frac{r^4}{s^4}+1\right)-\frac{2 r^4}{r^4+s^4}\right| \le \ln \left(\frac{r^4}{s^4}+1\right),
\end{equation}
and when $r>r_c$, it reduces to a trivial inequality
\begin{equation}
\frac{r^2}{r^4+s^4}\ge 0,
\end{equation}
and for $r<r_c$, it recovers the inequality of the NEC or the WEC
\begin{equation}
\frac{r^4}{r^4+s^4}\le \ln \left(\frac{r^4}{s^4}+1\right),
\end{equation}
which holds in all ranges of parameter space.

\begin{figure}[!ht]
	\centering
	\begin{subfigure}[b]{0.3\textwidth}
		\includegraphics[width=\textwidth]{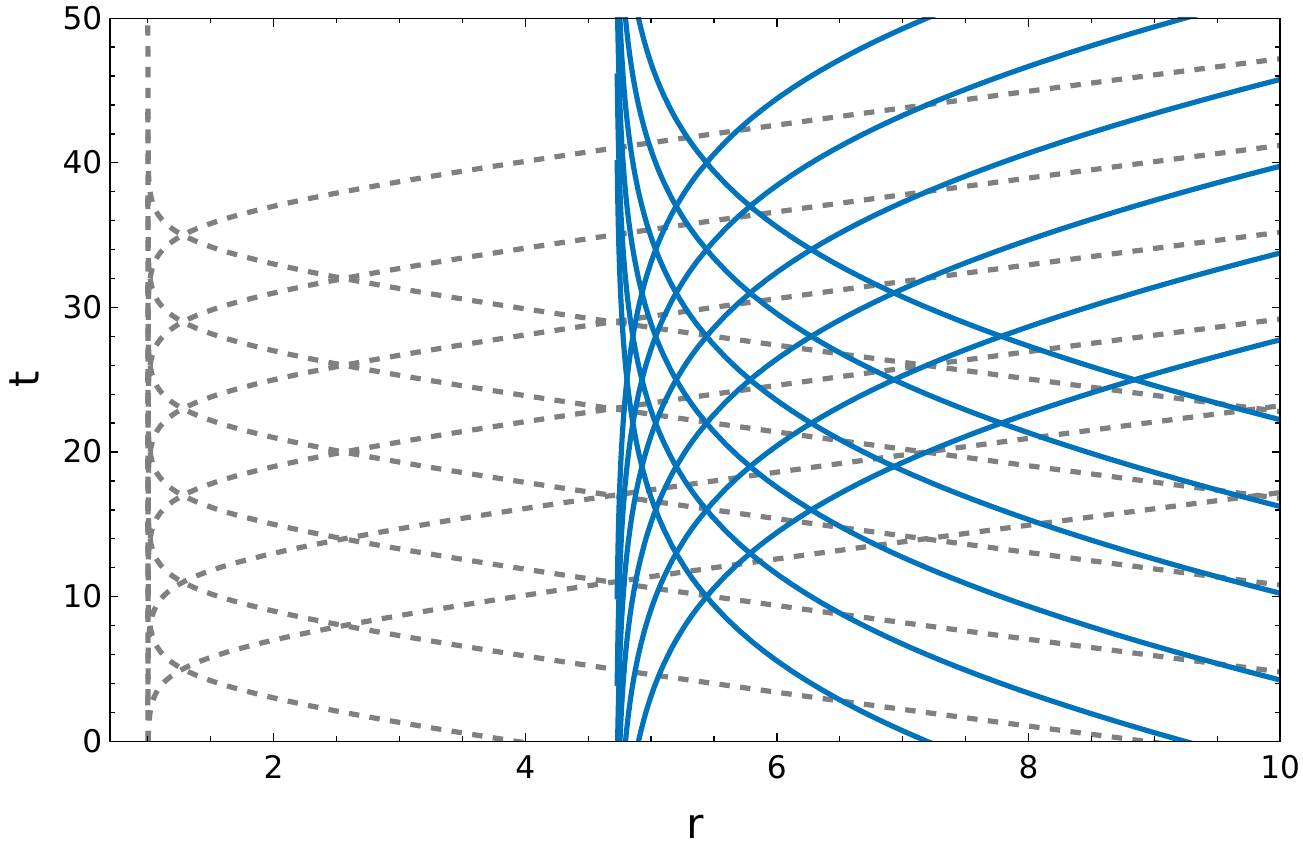}
		\caption{Gaussian-Ricci.}
		\label{fig:null_ricci_gauss}
	\end{subfigure}
	\begin{subfigure}[b]{0.3\textwidth}
		\includegraphics[width=\textwidth]{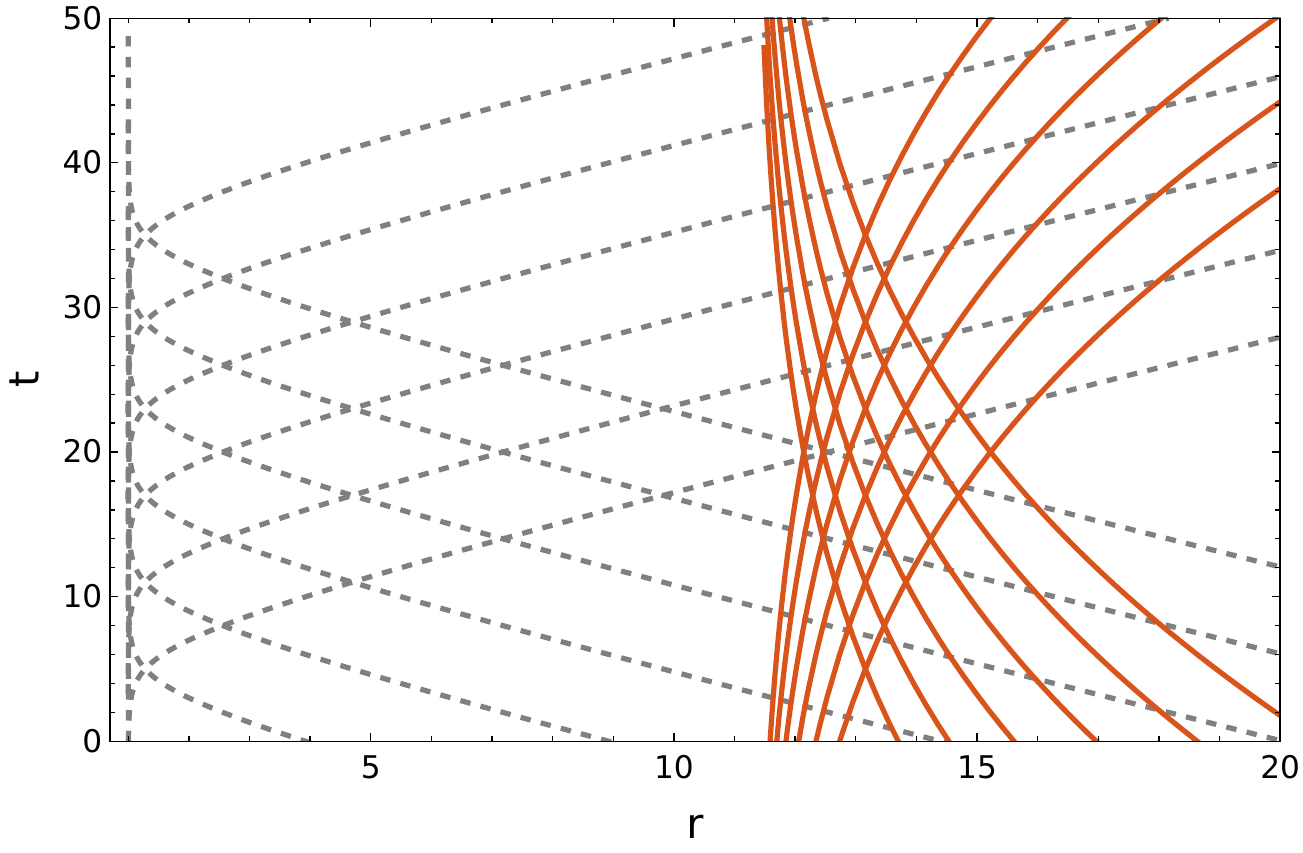}
		\caption{Sech-Ricci.}
		\label{fig:null_ricci_rech}
	\end{subfigure}
    \begin{subfigure}[b]{0.3\textwidth}
		\includegraphics[width=\textwidth]{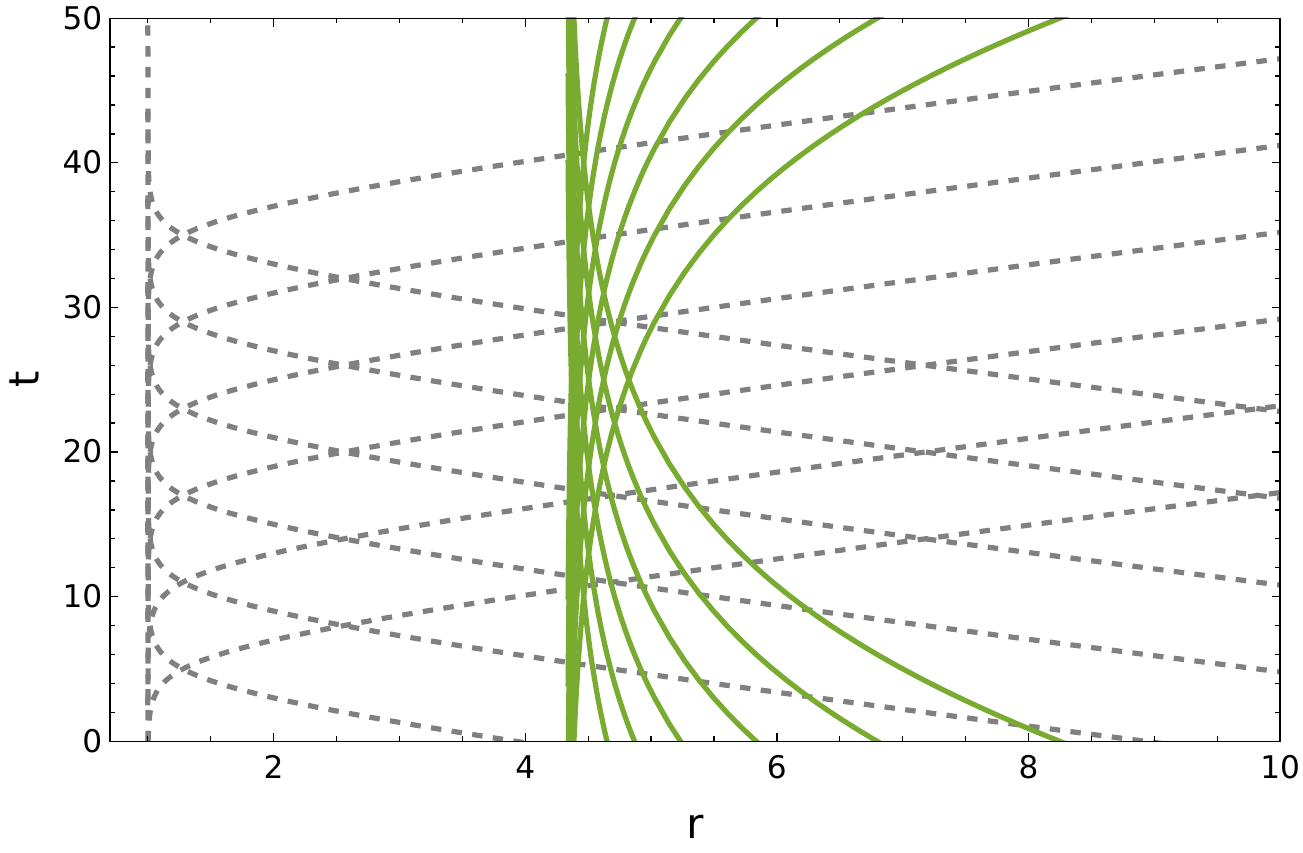}
		\caption{Fuzzy-Ricci.}
		\label{fig:null_ricci_fuzzy}
	\end{subfigure}
\captionsetup{width=.9\textwidth}
	\caption{Radial null geodesics for Ricci approach. The gray dashed lines represent the incoming (outgoing) radial null geodesics outside the event horizon of a Schwarzschild black hole. The blue, orange, and green lines correspond to the Gaussian-Ricci, Sech-Ricci, and Fuzzy-Ricci cases constructed via the Ricci approach, respectively.}
	\label{fig:null-Ricci}
\end{figure}

The radial null gedesics out side horizons for these tree cases are shown in Fig.\ \ref{fig:null-Ricci}.

\subsection{Combinations of bell-shaped functions} 
\label{sec:combination}

To construct regular black holes via the Weyl scalar, we use a combination of bell-shaped functions.
The reason is demonstrated in Sec.\ \ref{sec:weyl}, i.e., removing singularities demands that $\sigma(0)=0$.

\subsubsection*{Case one}

We propose the following form for $\sigma(r)$
\begin{equation}
    \sigma(r)=\frac{A r \left(p-\frac{r}{s}\right)}{1+\left(\frac{r}{s}\right)^6},
\end{equation}
where $p$ is a parameter to be determined by boundary conditions discussed below. The choice of powers in the numerator and denominator ensures the desired behavior of $\sigma(r)$ as $r \to 0$ and $r \to \infty$, as previously outlined in Sec.\ \ref{sec:weyl}. Additionally, the power of $r$ in the denominator is chosen to be even to avoid introducing singularities along the entire $r$-axis.

The two integrals appearing in Eq.\ \eqref{eq:sol-w} can be computed analytically
\begin{equation}
\begin{split}
    r^2 \int \sigma (r)\, \dif r 
    =\frac{A r^2 s^2 p}{12}  
    \Big[
    &\ln \frac{r^4-r^2 s^2+s^4}{(r^2+s^2)^2}
    +2 \left(\sqrt{3} +\frac{2}{p}\right) \arctan\left(\frac{2 r}{s}+\sqrt{3}\right)\\
    &+2 \left(\sqrt{3} -\frac{2}{p}\right) \arctan\left(\sqrt{3}-\frac{2 r}{s}\right)-\frac{4}{p} \arctan\left(\frac{r}{s}\right)
    \Big],
\end{split}    
\end{equation}
and
\begin{equation}
\begin{split}  
  r^3 \int \frac{\sigma (r)}{r}\, \dif r 
    &=
    -\frac{A r^3 s}{12}  \Big[2 \sqrt{3} p \arctanh\left(\frac{\sqrt{3} r s}{r^2+s^2}\right)+2 \left(p+\sqrt{3}\right) \arctan\left(\frac{2 r}{s}+\sqrt{3}\right)\\
    &+2 \left(\sqrt{3}-p\right) \arctan\left(\sqrt{3}-\frac{2 r}{s}\right)+4 p \arctan\left(\frac{r}{s}\right)+\ln \frac{r^4-r^2 s^2+s^4}{(r^2+s^2)^2}\Big].
\end{split}  
\end{equation}
Using these results, we find the asymptotic behavior of the mass function $m_+(r)$ from Eq.\ \eqref{eq:sol-w}. As $r \to 0$, we obtain
\begin{equation}
\label{eq:mp-0}
    m_+ =r^2 \left(-\frac{\pi  A p s^2}{3 \sqrt{3}}+c_1\right)+r^3 \left(-\frac{\pi  A s}{3 \sqrt{3}}+c_2\right)+O\left(r^4\right),
\end{equation}
and $r\to \infty$, 
\begin{equation}
\label{eq:mp-inf}
    m_+ =r^3 \left(-\frac{\pp  A p s}{3} +c_2\right)+r^2 \left(-\frac{\pp  A s^2}{6} +c_1\right)+\frac{A s^5}{12 r}+O\left(r^{-2}\right).
\end{equation}
The corresponding solution $m_-(r)$ can be obtained by replacing $A \to -A$, so in the following we focus only on $m_+(r)$.

The conditions for a regular black hole require that the $r^2$ term vanish as $r \to 0$ and that both $r^2$ and $r^3$ terms vanish as $r \to \infty$. Imposing these constraints fixes the constants
\begin{equation}
\label{eq:cond1}
    c_1= \frac{\pp  A s^2}{6} ,\quad
    c_2= \frac{\pp  A s}{2 \sqrt{3}},\quad
    p= \frac{\sqrt{3}}{2}.
\end{equation}
The Kretschmann scalar is finite along the entire $r$-axis. Near $r = 0$, it approaches 
\begin{equation}
    K = \frac{8}{9} \pp ^2 A^2 s^2+O\left(r\right),
\end{equation}
while as $r \to -\infty$,
\begin{equation}
    K=32 \pp ^2 A^2 s^2+\frac{32 \pp ^2 A^2 s^3}{\sqrt{3} r}+O\left(r^{-2}\right),
\end{equation}
and as $r \to \infty$,
\begin{equation}
    K=\frac{14 A^2 s^{10}}{9 r^8}+O\left(r^{-9}\right).
\end{equation}

\subsubsection*{Case two}

After introducing the previous model, we now propose an alternative construction for the function $\sigma(r)$, aiming for a different regular solution with similar desirable properties. Specifically, we consider
\begin{equation}
    \sigma(r)= \frac{A r }{s}\me^{-\frac{r^2}{s^2}} \left(p-\frac{r}{s}\right)
\end{equation}
where $p$ is a dimensionless parameter to be determined by boundary conditions. This form ensures that $\sigma(r)$ vanishes sufficiently fast at large distances and remains regular at the origin, thus satisfying the physical requirements for a smooth matter distribution.

With this choice of $\sigma(r)$, the mass functions $m_\pm(r)$ can be computed by direct integration, leading to
\begin{equation}
    m_\pm =
    c_2 r^3+c_1 r^2
    \mp\frac{A r^2}{4 s}   \left[2 \sqrt{\pp } p \frac{r}{s}  \text{erf}\left(\frac{r}{s}\right)+\sqrt{\pp }  \text{erf}\left(\frac{r}{s}\right)+2 p \me^{-\frac{r^2}{s^2}}\right]
\end{equation}
where $\text{erf}(x)$ is the error function. 
As before, our focus is primarily on the positive branch $m_+(r)$, with the negative branch obtainable via the replacement $A \to -A$.

To understand the local behavior of $m_+(r)$, we expand it in the limits $r \to 0$ and $r \to \infty$. Near the origin, we find
\begin{equation}
    m_+ = r^2 \left(-\frac{A p s}{2}+c_1\right)+r^3 \left(-\frac{A}{2}+c_2\right)+O\left(r^4\right),
\end{equation}
while at large distances the expansion yields
\begin{equation}
    m_+ = r^3 \left(-\frac{\sqrt{\pp } A p}{2} +c_2\right)+r^2 \left(-\frac{\sqrt{\pp } A s}{4} +c_1\right)+O\left(r^{-4}\right).
\end{equation}

To eliminate undesired terms and ensure asymptotic flatness, we require that the coefficients of $r^2$ vanish at the center and both the $r^2$ and $r^3$ terms vanish at infinity. These conditions uniquely fix the parameters as
\begin{equation}
\label{eq:cond2}
    p= \frac{\sqrt{\pp }}{2},\quad
    c_2= \frac{\pp  A}{4},\quad
    c_1= \frac{\sqrt{\pp } A s}{4}. 
\end{equation}

Having fully determined the parameters, we examine the behavior of the Kretschmann scalar $K$. 
Importantly, we find that $K$ remains finite everywhere, confirming the regular nature of the spacetime. Explicitly, the limiting values are
\begin{equation}
    \lim_{r\to \infty} K=0,\quad
    \lim_{r\to 0} K = 6A^2(\pp-2)^2,\quad
    \lim_{r\to-\infty} K = 24 A^2 \pp^2.
\end{equation}
Thus, this construction successfully yields a regular black hole solution, free from curvature singularities both at the center and spatial infinity.

\subsubsection*{Case three}

Encouraged by the success of the above constructions, we now present a third model based on a slightly modified functional form of $\sigma(r)$. In this case, we choose
\begin{equation}
    \sigma=\frac{A r }{s} \left(p-\frac{r}{s}\right) \ \text{sech}\left(\frac{r}{s}\right),
\end{equation}
where $\text{sech}(x)$ is the hyperbolic secant function. Compared to the previous Gaussian-based profiles, this choice leads to a different decay behavior at large distances while preserving regularity at the origin.

The integration of this $\sigma(r)$ yields the mass functions
\begin{equation}
    m_\pm =
    c_2 r^3+c_1 r^2\pm
    \frac{A r^2 (r-p s)}{2}  \me^{r/s}  \Phi \left(-\me^{\frac{2 r}{s}},2,\frac{1}{2}\right)\mp\frac{A r^2 s}{2}  \me^{r/s} \Phi \left(-\me^{\frac{2 r}{s}},3,\frac{1}{2}\right),
\end{equation}
where $\Phi(z, n, a)$ denotes the Lerch transcendent function. 

\begin{figure}[!ht]
	\centering
	\begin{subfigure}[b]{0.3\textwidth}
		\includegraphics[width=\textwidth]{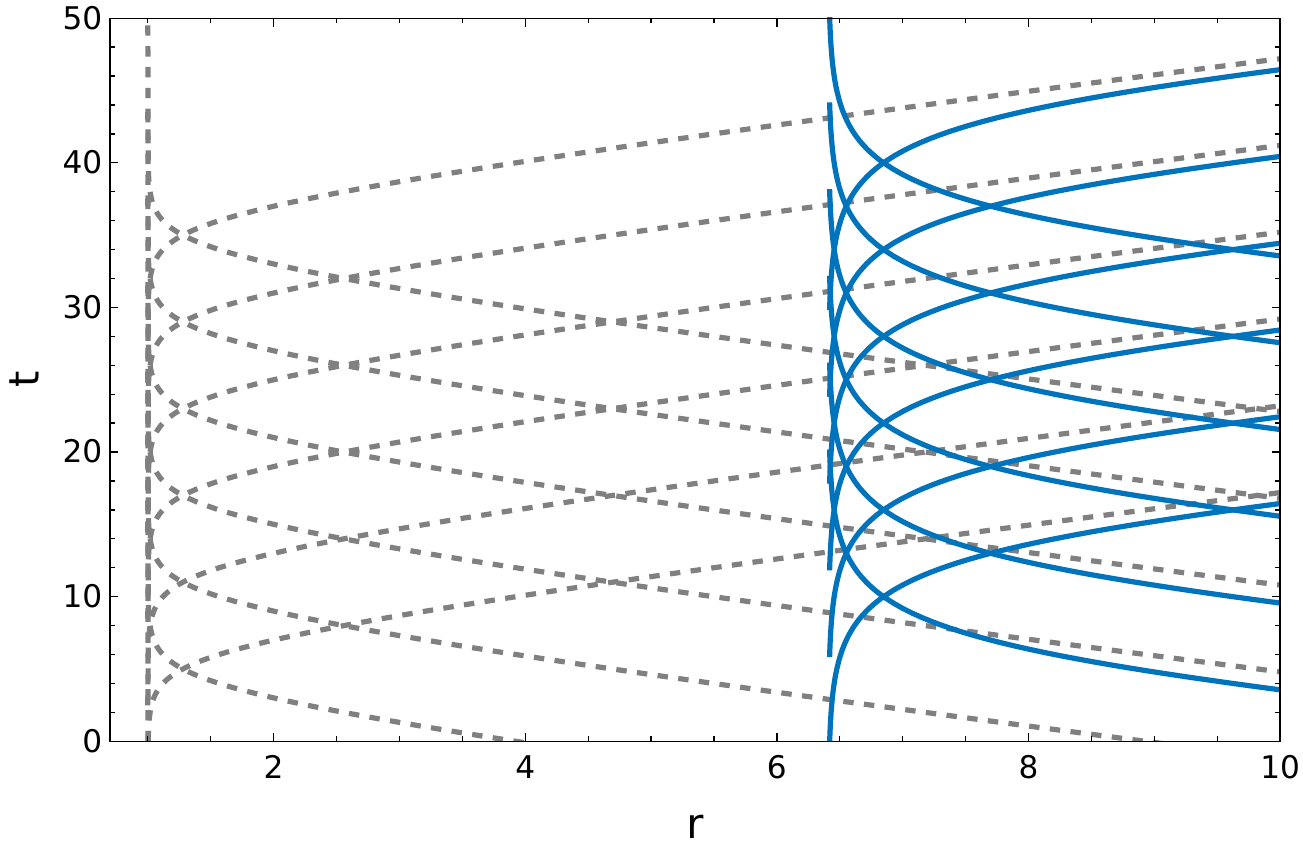}
		\caption{Null-Gaussian-Ricci.}
		\label{fig:null_weyl_gauss}
	\end{subfigure}
	\begin{subfigure}[b]{0.3\textwidth}
		\includegraphics[width=\textwidth]{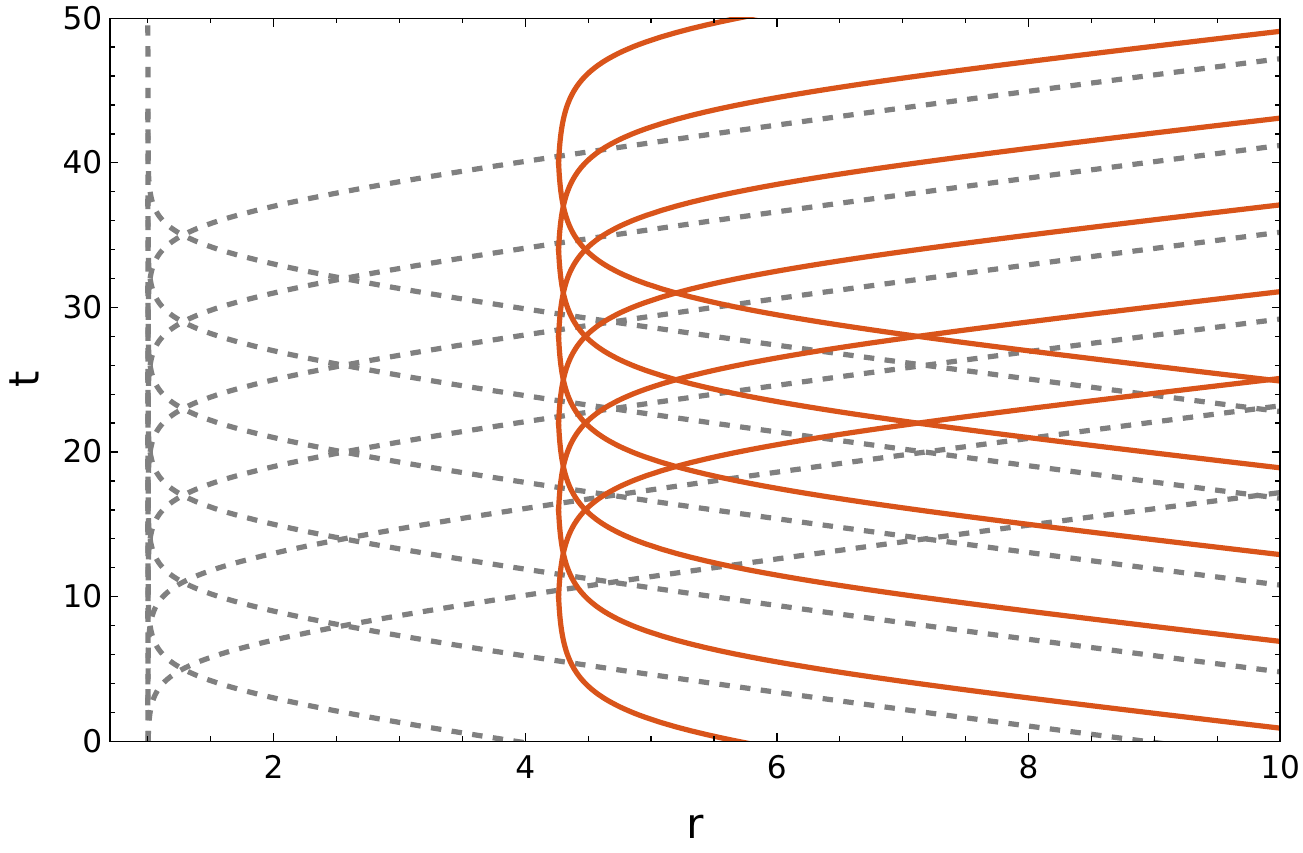}
		\caption{Null-Sech-weyl.}
		\label{fig:null_weyl_rech}
	\end{subfigure}
    \begin{subfigure}[b]{0.3\textwidth}
		\includegraphics[width=\textwidth]{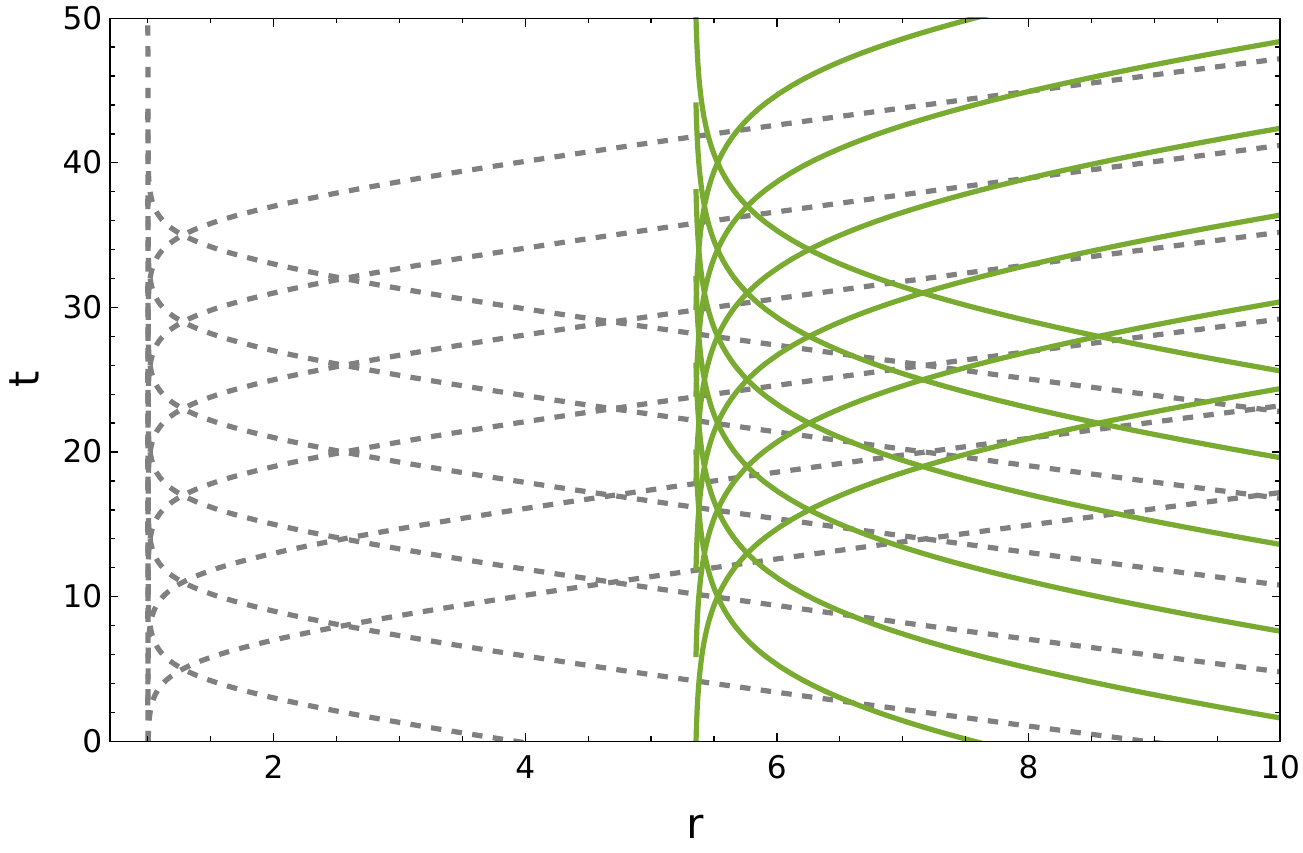}
		\caption{Null-Fuzzy-weyl.}
		\label{fig:null_weyl_fuzzy}
	\end{subfigure}
\captionsetup{width=.9\textwidth}
	\caption{Radial null geodesics for Weyl approach. The gray dashed lines represent the incoming (outgoing) radial null geodesics outside the event horizon of a Schwarzschild black hole. The blue, orange, and green lines correspond to the Gaussian-Weyl, Sech-Weyl, and Fuzzy-Weyl cases constructed via the Weyl approach, respectively.}
	\label{fig:null-weyl}
\end{figure}

Expanding $m(r)$ near the origin, we obtain
\begin{equation}
   m(r)= r^2 \left[-\frac{1}{8} A s \left(16 C p+\pi ^3\right)+c_1\right]+r^3 \left[-\frac{1}{2} A (4 C+\pi  p)+c_2\right]+O\left(r^4\right),
\end{equation}
where $C$ is a Catalan number arising from the series expansion of the Lerch transcendent near $r=0$.
At large distances, the asymptotic form of the mass function reads
\begin{equation}
    m(r) = r^3 (-\pp  A p+c_2)-\frac{1}{4} r^2 \left(\pp ^3 A s-4 c_1\right)+O\left(r^4\right).
\end{equation}
Requiring regularity at the center and asymptotic flatness at infinity fixes the parameters as
\begin{equation}
\label{eq:cond3}
    p= \frac{\pp ^3}{16 C},\quad
    c_1= \frac{1}{4} \pp ^3 A s,\quad
    c_2= \frac{\pp ^4 A}{16 C}.
\end{equation}
Thus, this third construction offers yet another class of regular solutions, demonstrating the flexibility of our method in designing regular black holes.
The Fig.\ \ref{fig:null-weyl} exibits the outgoing and incoming radial null geodesics outside the horizons.

\begin{figure}[!ht]
\centering
\includegraphics[width=0.45\textwidth]{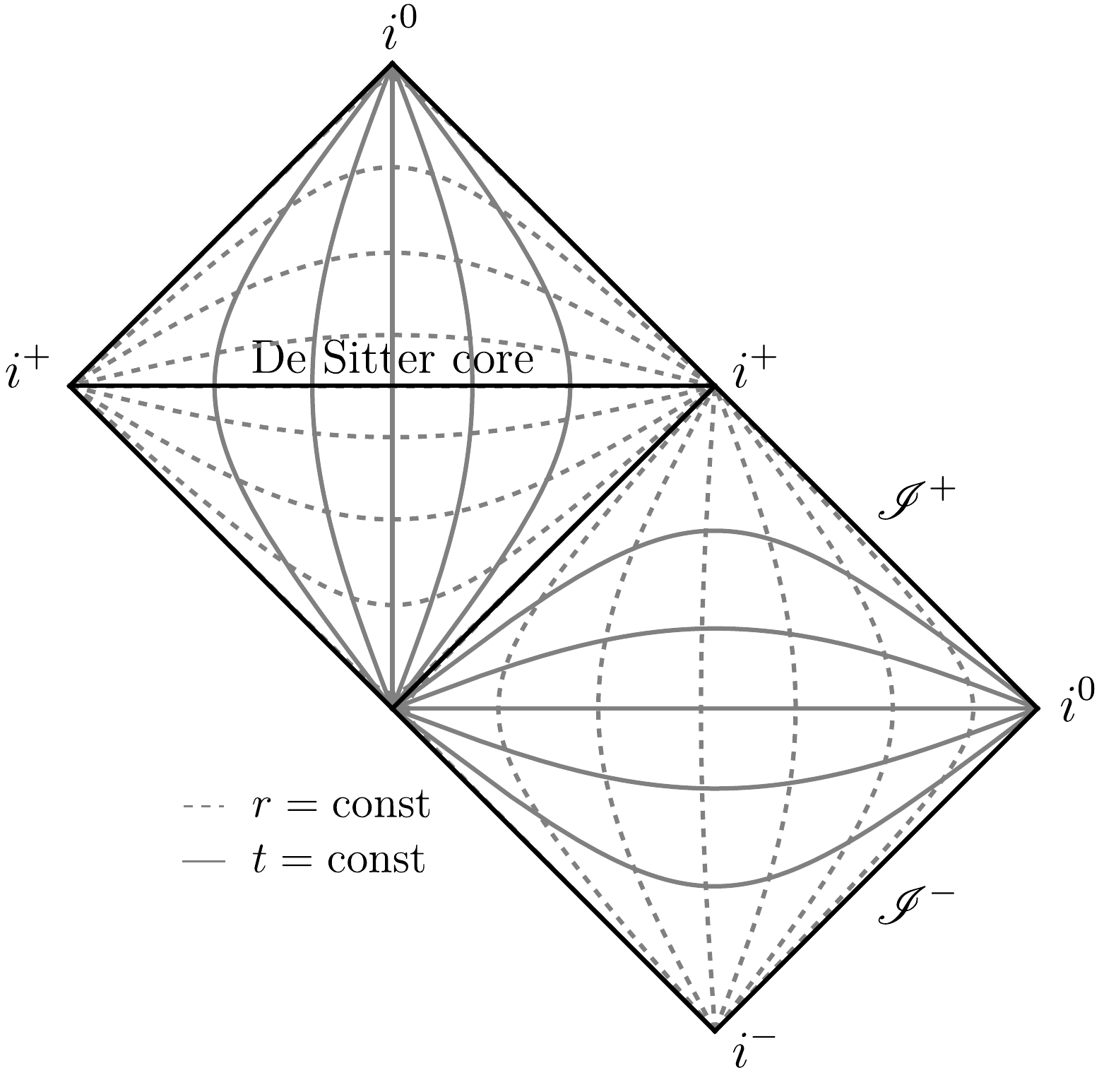}
\captionsetup{width=0.9\textwidth}
\caption{Penrose diagrams for all six modes.}
\label{fig:penrose_dig}
\end{figure}

The choice of bell-shaped profiles for $\beta(r)$ and $\sigma(r)$ is motivated by both physical intuition and mathematical convenience. Physically, these functions represent localized quantum or modified gravity effects that peak near the black hole core ($r\sim 0$) and decay rapidly at larger distances, ensuring classical GR recovery asymptotically. Gaussian forms (e.g., $e^{-r^2}$) provide sharp localization, mimicking short-range quantum corrections like those in asymptotic safety. Hyperbolic secant (sech) profiles offer broader tails, suitable for modeling extended de Sitter cores from vacuum energy. Rational forms (e.g., $1/(1+r^4)$) balance analyticity and slow decay, akin to nonlinear electrodynamics. Mathematically, they are smooth, singularity-free, and satisfy integrability conditions, allowing tunable horizons while respecting energy conditions in outer regions. For the six kinds of black holes constructed by the method proposed in this paper, they all possess an event horizon, and the curvature K has no singularities on the entire real axis (especially on the negative real axis). Therefore, they share identical Penrose diagrams, see Fig.\ \ref{fig:penrose_dig}.

\section{Analysis of quasinormal modes}
\label{sec:QNMs}

In this section, we apply the QNM analysis to study the models constructed above. 
Quasinormal modes are particularly important in the context of modified gravity, 
as they provide a dynamical diagnostic for the stability of black hole solutions 
and may encode imprints of the underlying curvature regularization.  

In Einstein gravity, the perturbation equation can be cast into a Schr\"odinger-like form,
\begin{equation}   
 \label{eq:wave_equation} 
 \left(-\partial^2_{r^*}+V_{\mathrm{eff}}\right)\phi(r^*)=\omega^2 \phi(r^*),  
\end{equation}
where $r^*$ is the tortoise coordinate defined by
\begin{equation}
\label{eq:tortoise}
    \frac{\dif r^*}{\dif r}=\frac{1}{f(r)},
\end{equation}
and $V_\mathrm{eff}$ is the effective potential given by (see App.\ \ref{app:axial} for axial perturbations)
\begin{equation}
    \label{eq:Veff}
    V_\mathrm{eff}=f(r) \left[\frac{l(l+1)}{r^2}+\frac{1-\mathfrak{s}^2}{r}\frac{\dif f(r)}{\dif r}\right].
\end{equation}
Here $\mathfrak{s}=0,1,2$ denotes the spin of the perturbation. In what follows, we focus exclusively on gravitational perturbations with $\mathfrak{s}=2$.  

\begin{figure}[!ht]
	\centering
	\begin{subfigure}[b]{0.45\textwidth}
		\includegraphics[width=\textwidth]{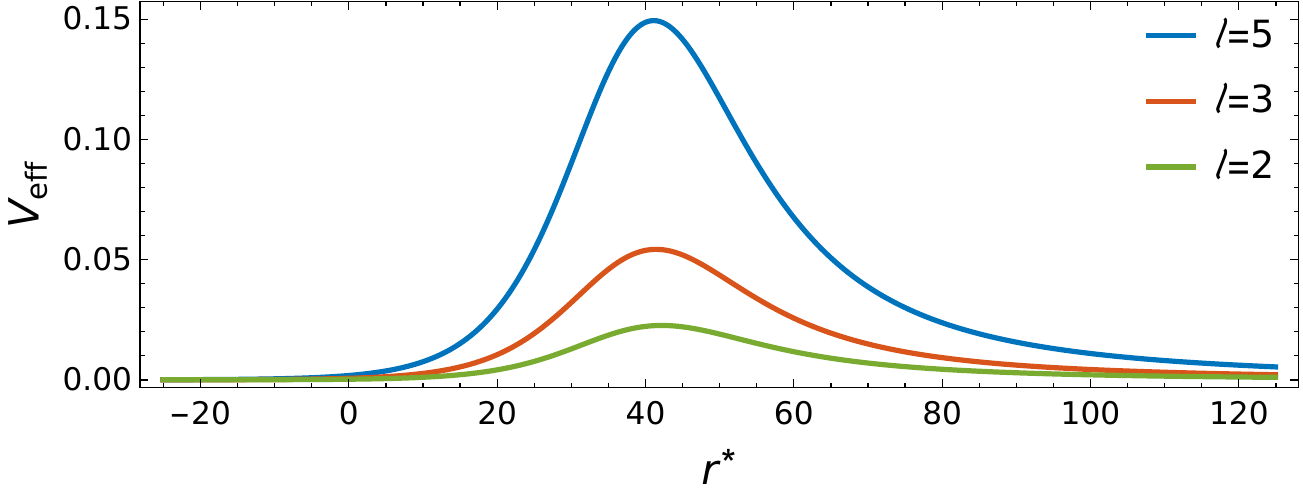}
		\caption{Gaussian-Ricci.}
		\label{fig:Veff-gauss}
	\end{subfigure}
	\begin{subfigure}[b]{0.45\textwidth}
		\includegraphics[width=\textwidth]{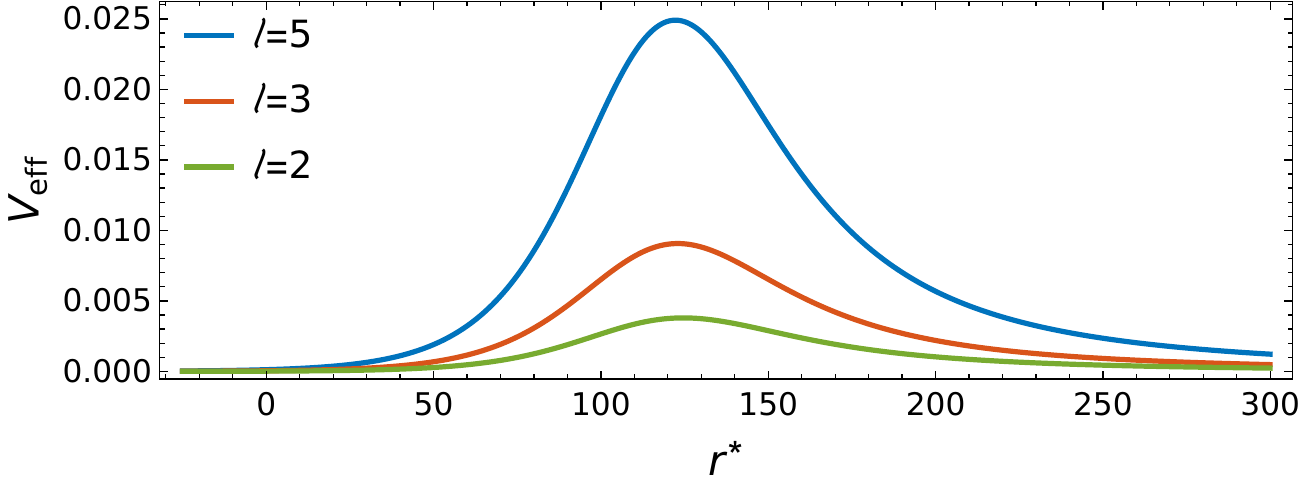}
		\caption{Sech-Ricci.}
		\label{fig:Veff-sech}
	\end{subfigure}
    \begin{subfigure}[b]{0.45\textwidth}
		\includegraphics[width=\textwidth]{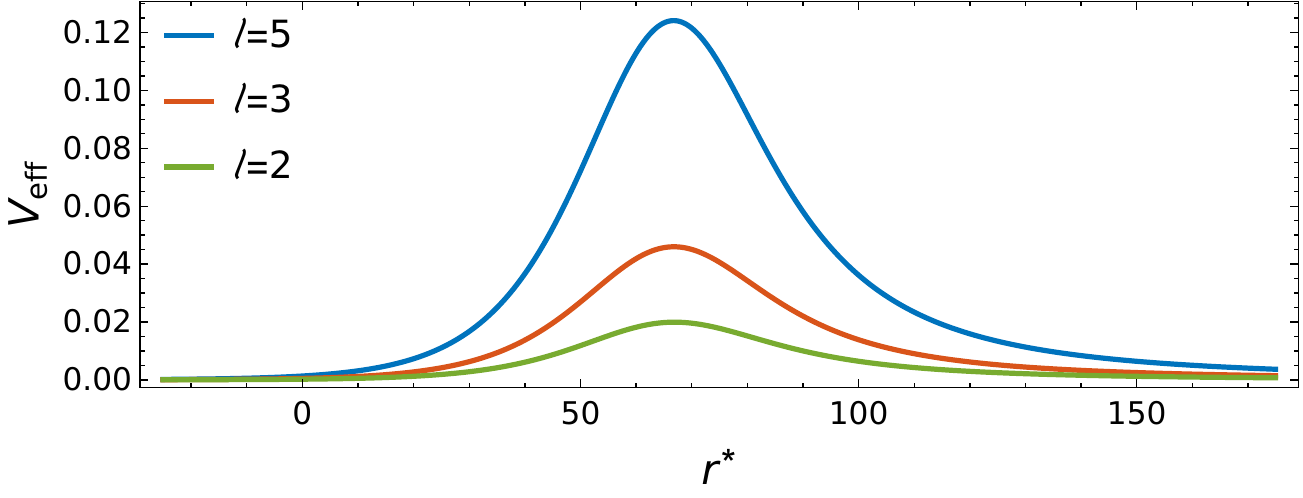}
		\caption{Fuzzy-Ricci.}
		\label{fig:Veff-fuzzy}
	\end{subfigure}
\captionsetup{width=.9\textwidth}
	\caption{Effective potentials for models constructed by the Ricci-scalar approach.}
	\label{fig:Veff-Ricci}
\end{figure}

To evaluate the QNMs, we adopt the finite difference method, as in 
our recent work~\cite{Zhang:2025dzt}. For all Ricci-scalar models 
we fix $r_0 = 0$ by symmetry and set $A = 7$, $s = 1$; for all 
Weyl-scalar models the parameters $p$ and $c_i$ are determined by 
the boundary conditions (Eqs.~\eqref{eq:cond1}, \eqref{eq:cond2}, \eqref{eq:cond3}), and we likewise 
adopt $A = 7$, $s = 1$. These values are chosen to place each 
solution within the single-horizon regime (cf.\ Figs.~\ref{fig:horizon-gauss}, \ref{fig:horizon-sech}, \ref{fig:horizon-fuzzy}) 
while satisfying the NEC, WEC, and DEC established in Sec.~\ref{sec:models}. 
We begin with the models constructed from the Ricci scalar, whose 
effective potentials are displayed in Fig.~\ref{fig:Veff-Ricci}.

For $\ell=2,3,5$, all three Ricci-scalar models exhibit the familiar single-barrier structure, 
but with clear differences in both height and width. 
The Gaussian-Ricci case produces the steepest and narrowest peak, whose height grows rapidly with $\ell$. 
In contrast, the Sech-Ricci potential is broader and shallower, especially at small $\ell$, 
while the Fuzzy-Ricci potential takes an intermediate form.  The quasinormal frequencies for the fundamental overtone $n=0$ are presented in Table \ref{Tab:qnm_ricci}.
Here we use the Prony method\cite{Berti:2009kk} to extract quasinormal frequencies from the corresponding waveform.

\begin{table}[!ht]
    \centering
    \caption{Quasinormal frequencies for Ricci approach.}
 \begin{tabular}{ccccc}\hline
  
    $\ell$      & Schw & Gauss-Ricci   & Sech-Ricci  & Fuzzy-Ricci            \\ \hline
  
    $2$    &  0.74735 - 0.17793 i  & 0.14258 - 0.02892 i   &   0.05827 - 0.01074 i &  0.13917 - 0.02144 i \\ \hline
    $3$    &  1.19889 - 0.185406 i  & 0.23021 - 0.03032 i   &   0.09844 - 0.01146 i &  0.21201 - 0.02236 i \\ \hline
    $5$    &  2.02459 - 0.189741 i  & 0.43235 - 0.03081 i   &   0.15671 - 0.01217 i &  0.34723 - 0.02266 i \\ \hline
  \end{tabular}
    \label{Tab:qnm_ricci}
\end{table}

\begin{figure}[!ht]
\centering
	\begin{subfigure}[b]{0.45\textwidth}
		\includegraphics[width=\textwidth]{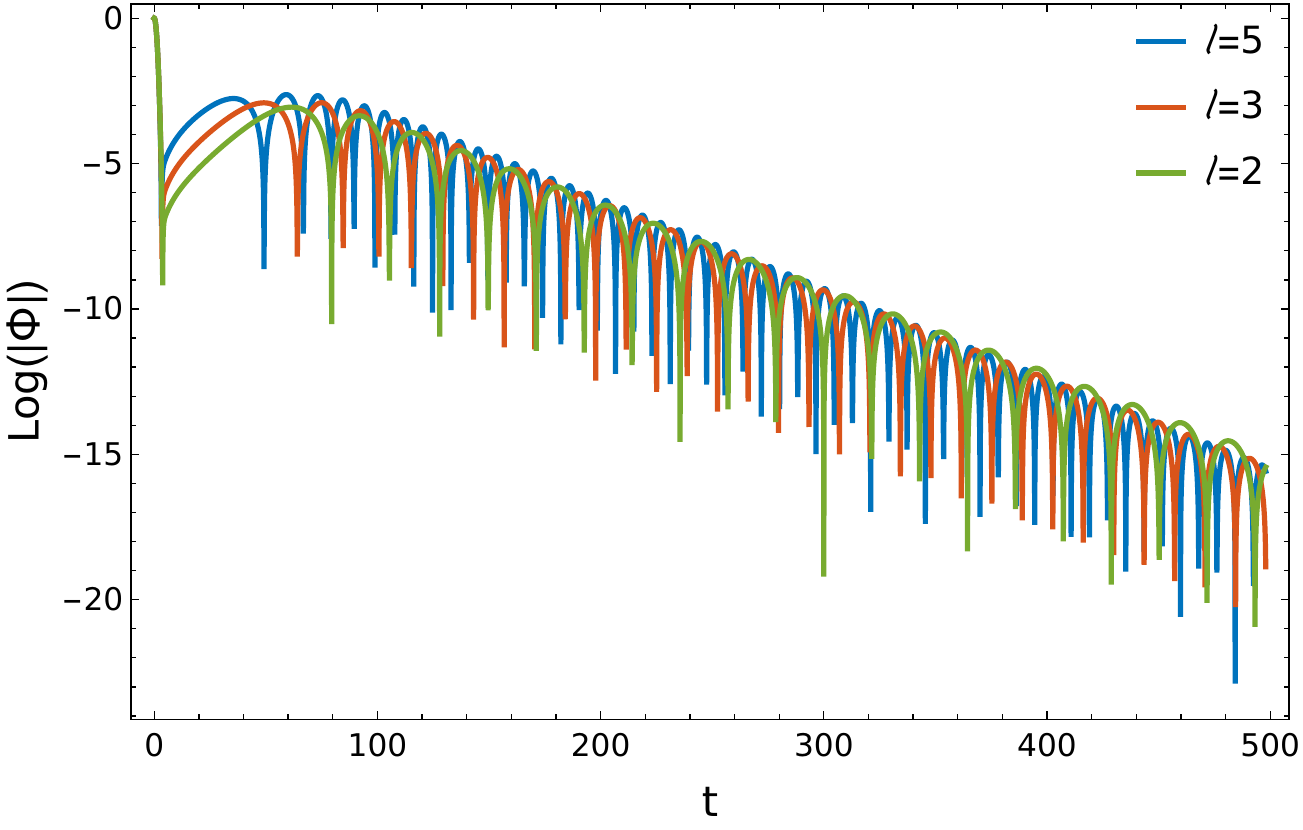}
		\caption{Gaussian-Ricci.}
	\end{subfigure}
    \begin{subfigure}[b]{0.45\textwidth}
		\includegraphics[width=\textwidth]{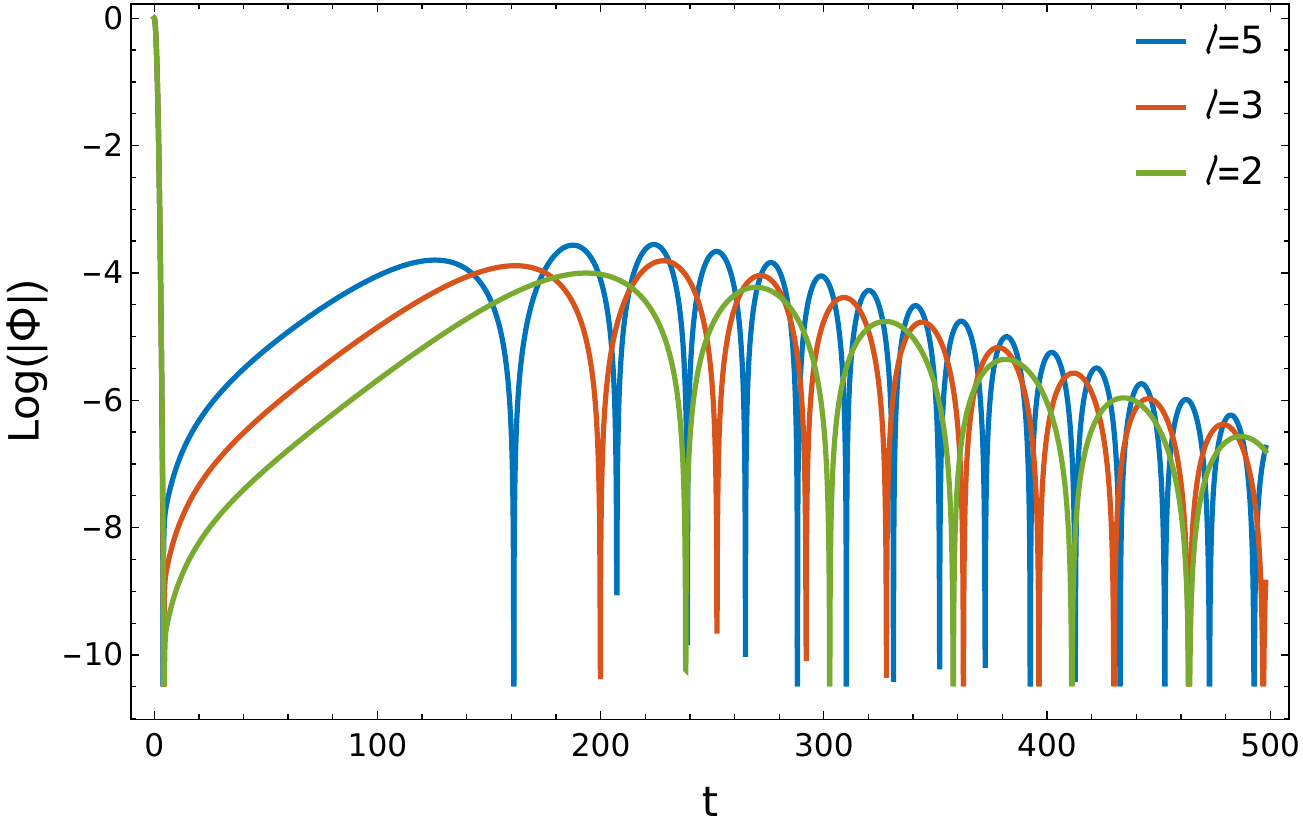}      
		\caption{Sech-Ricci.}
	\end{subfigure}
    \begin{subfigure}[b]{0.45\textwidth}
        \includegraphics[width=\textwidth]{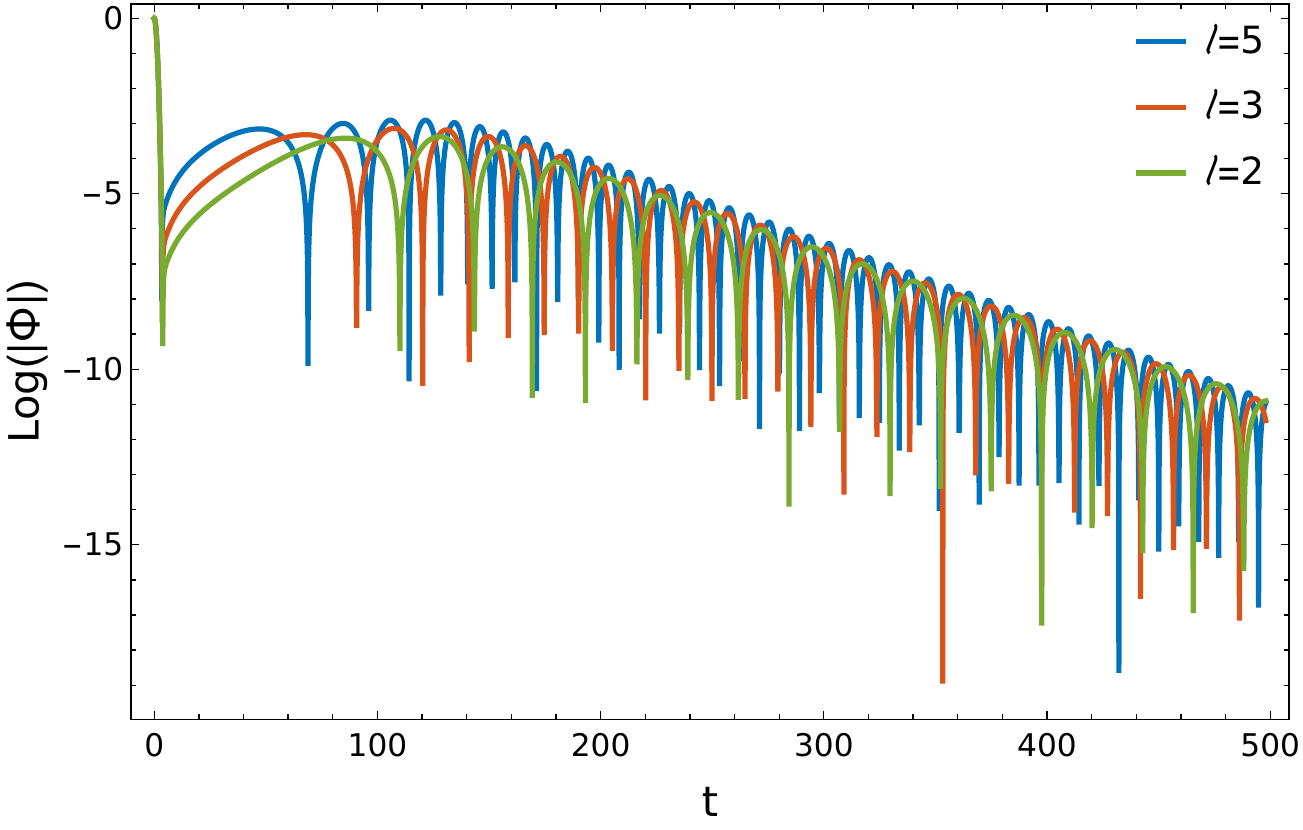}    
		\caption{Fuzzy-Ricci.}
	\end{subfigure}    
\captionsetup{width=0.9\textwidth}
\caption{Quasinormal waveforms of models constructed by the Ricci-scalar approach.}
\label{fig:qnm-ricci-2}
\end{figure}

The corresponding waveforms are shown in Fig.~\ref{fig:qnm-ricci-2}.  
The sharp Gaussian-Ricci barrier results in higher oscillation frequencies and more rapid damping, 
reflecting efficient energy leakage from the near-horizon region. 
The Sech-Ricci barrier, being broader and shallower, traps perturbations longer and leads to more persistent oscillations. 
The Fuzzy-Ricci case again interpolates between these two behaviors. 
As $\ell$ increases, all three models display the expected trend of higher frequencies and shorter damping times, 
yet their relative ordering remains robust.  

Several features of Table~\ref{Tab:qnm_ricci} deserve explicit comment. 
First, the QNM frequencies of all three regular models differ substantially 
from those of the Schwarzschild black hole—by well over $10\%$ in both 
the real and imaginary parts. This large discrepancy has a clear physical 
origin: unlike the Schwarzschild solution, which is characterized solely 
by its ADM mass $M$, our regular models possess additional free parameters 
($A$, $s$) that independently control the height and width of the effective 
potential. Even at fixed ADM mass, varying these parameters modifies the 
potential profile significantly and shifts the entire QNM spectrum. The 
deviations seen in Table~\ref{Tab:qnm_ricci} therefore reflect a genuine 
geometrical difference between the regular and singular spacetimes, not an 
inconsistency of the model. Indeed, by continuously tuning $A$ and $s$ 
toward the Schwarzschild limit (i.e., by making the curvature profile 
increasingly narrow and concentrated), the QNM frequencies can be made to 
approach those of Schwarzschild arbitrarily closely.

Second, the absolute values of the imaginary parts, $|\mathrm{Im}(\omega)|$, 
are systematically smaller for all regular models than for Schwarzschild at 
the same $\ell$. Since the damping timescale is $\tau = 1/|\mathrm{Im}(\omega)|$, 
a smaller $|\mathrm{Im}(\omega)|$ implies a \emph{longer} decay time, i.e., 
perturbations ring down more slowly. Crucially, this does not indicate 
reduced stability: the definitive criterion for linear stability under axial 
perturbations is $\mathrm{Im}(\omega) < 0$, which is satisfied for all modes 
in Table~\ref{Tab:qnm_ricci}. The slower decay is instead a direct 
consequence of the broader and shallower effective potentials produced by 
the bell-shaped curvature profiles. From an observational standpoint, it 
implies a longer-lived ringdown signal compared to a Schwarzschild black 
hole of the same ADM mass—a potentially distinctive signature of the 
regular interior that could, in principle, be probed by future 
precision gravitational-wave measurements.

Third, note that $|\mathrm{Im}(\omega)|$ increases only mildly with $\ell$ 
for the regular models. For instance, in the Gauss-Ricci case it rises 
from $0.02892$ at $\ell = 2$ to $0.03081$ at $\ell = 5$, whereas for 
Schwarzschild the corresponding change is from $0.17793$ to $0.18974$, 
already near saturation. This much flatter $\ell$-dependence of the damping 
rate constitutes an additional spectral feature characteristic of the 
regular geometry.

Physically, the Ricci-scalar analysis illustrates how the near-horizon geometry shapes the dynamical response of the black hole. 
The Gaussian-Ricci model corresponds to a ``stiffer'' configuration with rapid damping, 
while the Sech-Ricci potential resembles a ``softer'' resonant cavity supporting longer-lived oscillations. 
The Fuzzy-Ricci case provides an intermediate picture. 
This underscores that not only the barrier height but also its width and shape crucially determine the QNM spectrum.  

\medskip

\begin{figure}[!ht]
	\centering
	\begin{subfigure}[b]{0.45\textwidth}
		\includegraphics[width=\textwidth]{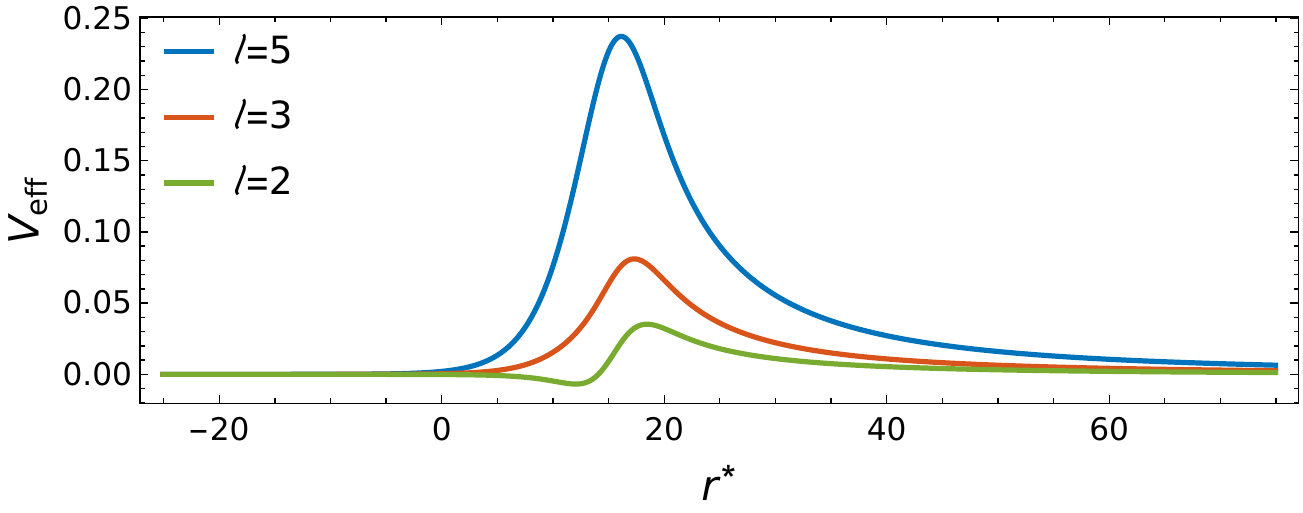}
		\caption{Gaussian-Weyl.}
        \label{fig:Gaussian-Weyl}
	\end{subfigure}
	\begin{subfigure}[b]{0.45\textwidth}
        \includegraphics[width=\textwidth]{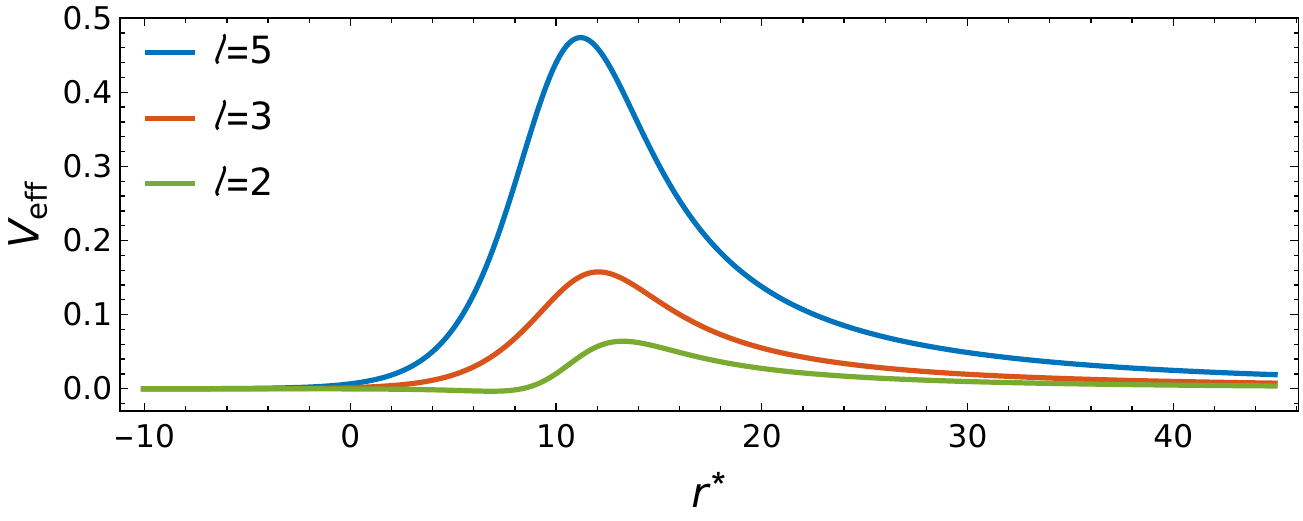}
		\caption{Sech-Weyl.}
	\end{subfigure}
    \begin{subfigure}[b]{0.45\textwidth}
        \includegraphics[width=\textwidth]{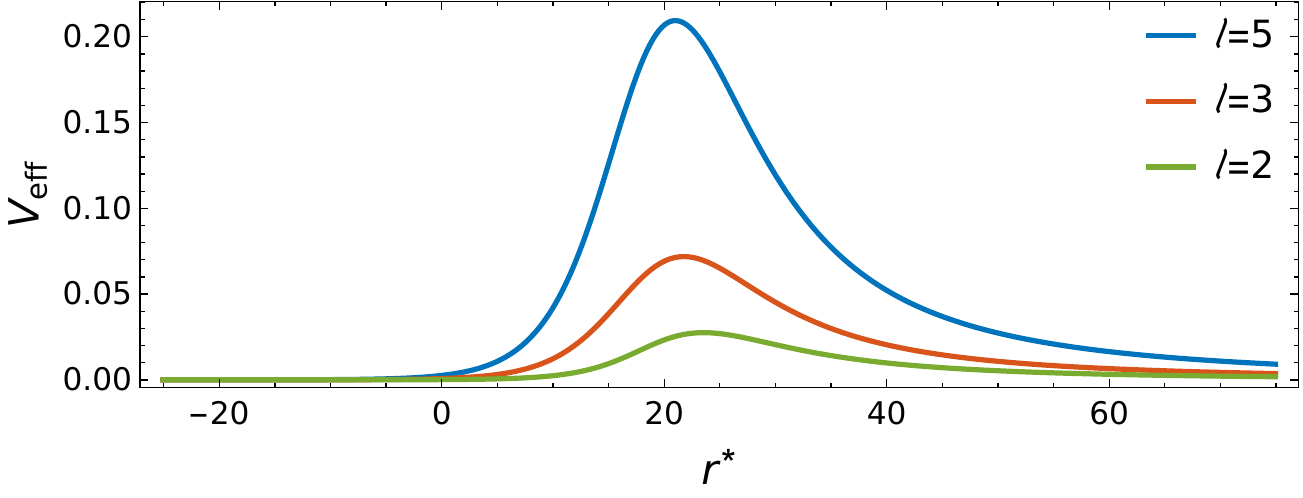}
		\caption{Fuzzy-Weyl.}
	\end{subfigure}
\captionsetup{width=.9\textwidth}
	\caption{Effective potentials for models constructed by the Weyl-scalar approach.}
	\label{fig:Veff-Weyl}
\end{figure}

\begin{figure}[!ht]
\centering
	\begin{subfigure}[b]{0.45\textwidth}
		\includegraphics[width=\textwidth]{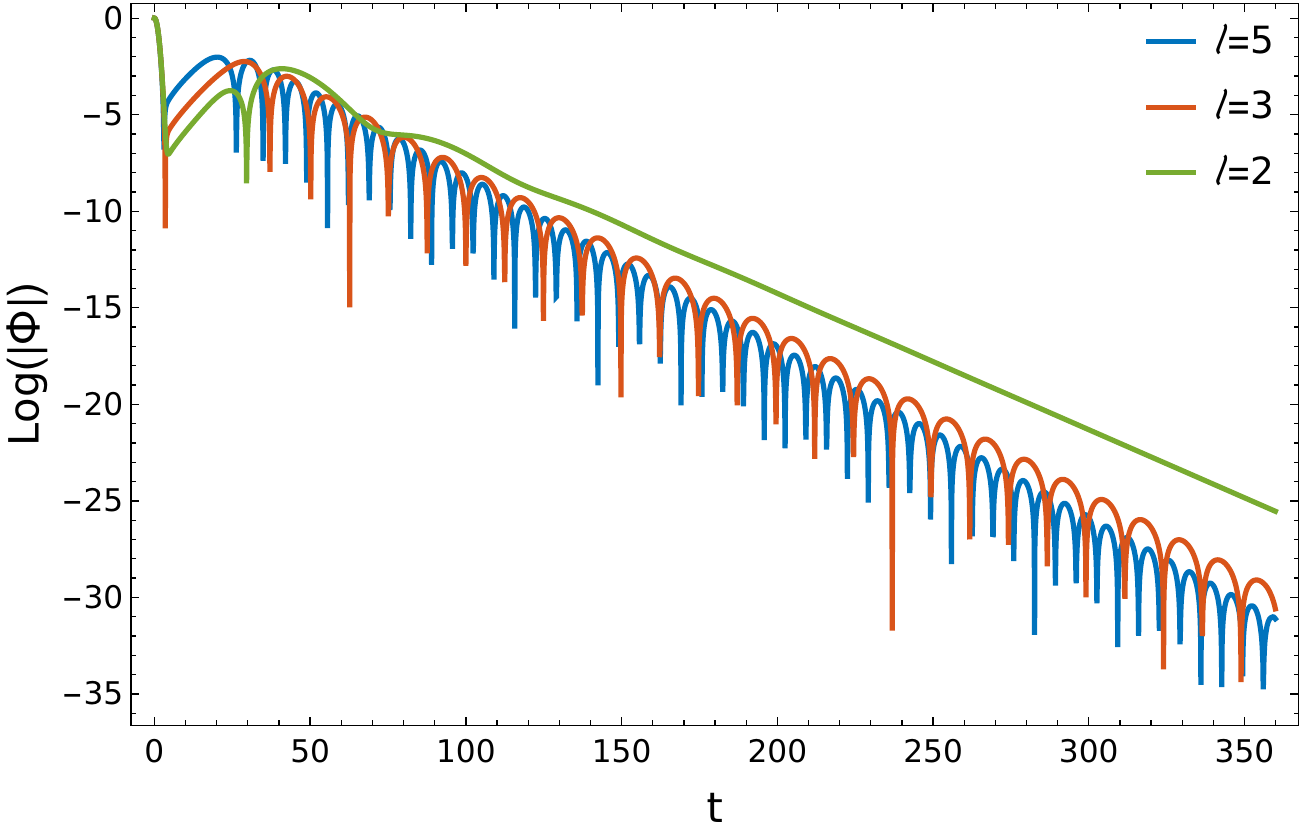}
		\caption{Gaussian-Weyl.}
	\end{subfigure}
    \begin{subfigure}[b]{0.45\textwidth}
		\includegraphics[width=\textwidth]{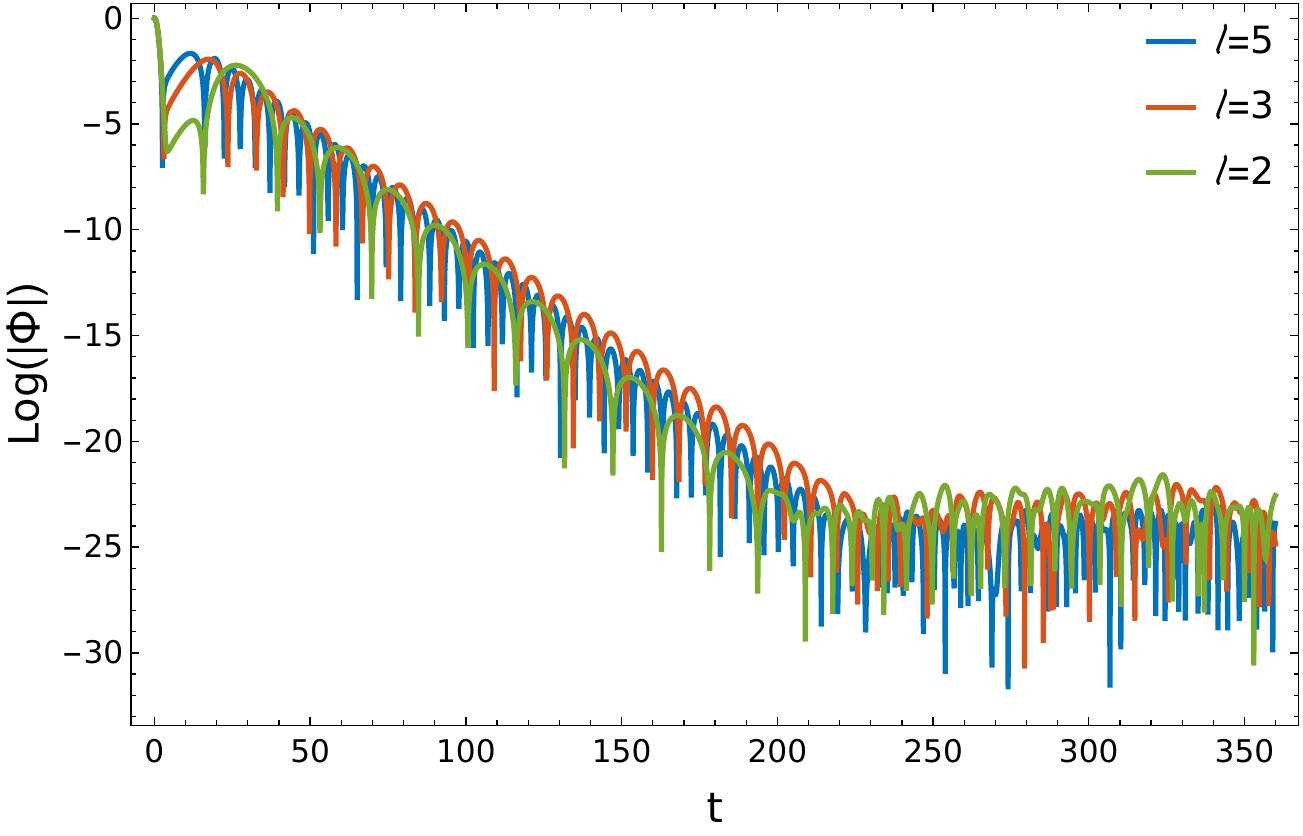}
		\caption{Sech-Weyl.}
	\end{subfigure}
    \begin{subfigure}[b]{0.45\textwidth}
		\includegraphics[width=\textwidth]{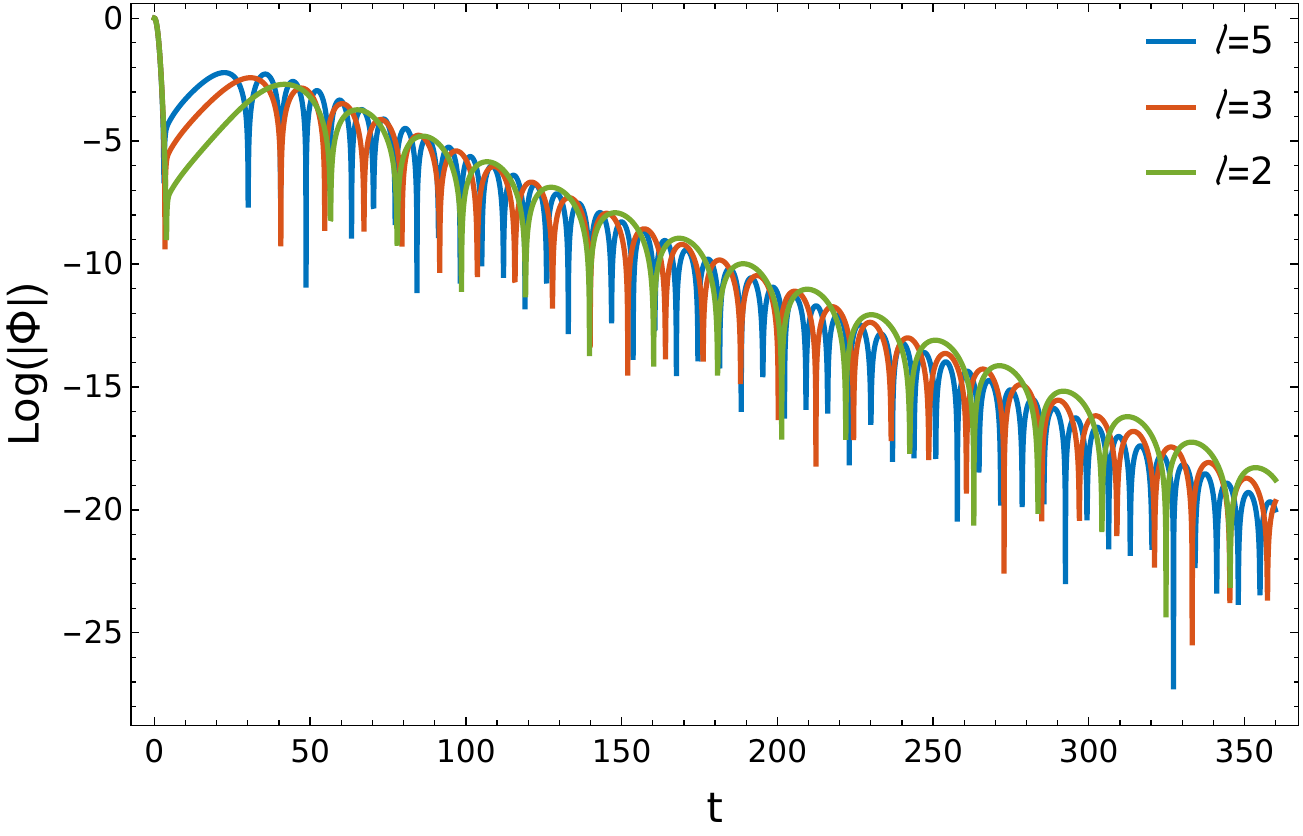}
		\caption{Fuzzy-Weyl.}
	\end{subfigure}    
\captionsetup{width=0.9\textwidth}
\caption{Quasinormal waveforms of models constructed by the Weyl-scalar approach.}
\label{fig:qnm-weyl}
\end{figure}

We now turn to the models constructed from the Weyl scalar, shown in Figs.~\ref{fig:Veff-Weyl} and~\ref{fig:qnm-weyl}.  
Like the Ricci-scalar constructions, the Weyl-scalar models share the single-barrier feature but differ in detail. 
Notably, the Gaussian-Weyl and Sech-Weyl cases develop secondary minima (valleys) for low $\ell$, 
while the Fuzzy-Weyl model retains a smoother profile without pronounced valleys. 
We also extract the fundamental quasinormal frequencies ($n=0$), which are summarized in Table \ref{Tab:qnm_weyl}.
\begin{table}[!ht]
    \centering
    \caption{Quasinormal frequencies for Weyl approach.The locations marked with an asterisk indicate instability.}
 \begin{tabular}{cccc}\hline
  
    $\ell$  & Gauss-Weyl   & Sech-Weyl  & Fuzzy-Weyl            \\ \hline
  
    $2$     & *                    &   0.20423 - 0.11662 i &   0.15293 - 0.05082 i\\ \hline
    $3$     & 0.25270 - 0.08376 i  &   0.36123 - 0.10347 i &   0.25825 - 0.05283 i\\ \hline
    $5$     & 0.44398 - 0.08841 i  &   0.64721 - 0.10920 I &   0.45212 - 0.05478 i\\ \hline
  \end{tabular}
    \label{Tab:qnm_weyl}
\end{table}

The waveforms exhibit clear signatures of these structural differences.
For the Gaussian-Weyl potential, the $\ell=2$ valley is relatively shallow, with a peak-to-valley ratio of $|V_p/V_v|=5.16795$.
The corresponding waveform resembles the higher-$\ell$ cases, showing standard damping but lacking sustained oscillations.
By contrast, the Sech-Weyl potential possesses a much deeper valley, with a significantly larger peak-to-valley ratio of $|V_p/V_v|=19.5718$.
In this case, the $\ell=2$ waveform develops noticeably longer-lived oscillations, which indicates that perturbations are partially trapped in the valley region before leaking out.
This comparison demonstrates that it is not merely the presence of a valley, but rather the relative depth of the valley compared to the peak height, that governs whether additional trapping becomes dynamically relevant.
It is also worth comparing the imaginary parts in 
Table~\ref{Tab:qnm_weyl} across the three Weyl-scalar models. 
The Sech-Weyl model has the largest $|\mathrm{Im}(\omega)|$ among 
the three, a trend that is consistent with its deeper potential 
valley and larger peak-to-valley ratio $|V_p/V_v| = 19.57$ at 
$\ell = 2$: the strong quasi-trapping promotes a more efficient 
exchange of energy between trapped and outgoing sectors, resulting 
in faster effective damping at higher $\ell$. By contrast, the 
Fuzzy-Weyl model, whose potential is smoother and lacks pronounced 
valleys, exhibits the smallest $|\mathrm{Im}(\omega)|$ and hence 
the slowest ringdown among the Weyl-scalar constructions. As with 
the Ricci-scalar models, all non-asterisked entries have 
$\mathrm{Im}(\omega) < 0$, confirming linear stability. The 
instability of the Gauss-Weyl $\ell = 2$ mode is associated with 
the negative-potential region visible in Fig.~\ref{fig:Gaussian-Weyl} at low $\ell$; 
for $\ell \geq 3$ the potential becomes everywhere non-negative 
and the corresponding QNMs are stable.
The Fuzzy-Weyl case once again exhibits intermediate behavior, with damping rates lying between these two extremes.

Taken together, the Ricci- and Weyl-scalar analyses highlight the sensitivity of QNMs to the fine structure of the effective potential. 
The Ricci-scalar models primarily illustrate the role of barrier height and width, 
while the Weyl-scalar models reveal that additional features such as valleys can qualitatively alter the low-$\ell$ dynamics. 
From a physical standpoint, Gaussian-type models behave like ``stiffer'' geometries with rapid damping, 
whereas Sech-type models act as ``softer'' configurations that sustain oscillations longer. 
Moreover, the emergence of valley-induced trapping in some Weyl-scalar cases suggests the possibility of quasi-resonant states or echo-like signatures.  

In summary,  
our analysis shows that quasinormal modes serve as a sensitive probe of the near-horizon structure of regular black holes. 
Different curvature-based constructions—Ricci versus Weyl—lead to distinctive effective potentials, 
which in turn imprint themselves on the QNM spectra. 
Although such fine features may remain beyond the reach of current detectors, 
future precision measurements of ringdown signals could potentially discriminate between these models, 
thereby providing an observational window into the nature of curvature regularization.

\section{Conclusion}
\label{sec:conclusion}

In this work, we propose a systematic method for constructing regular black holes by prescribing finite curvature invariants, 
specifically the Ricci scalar and Weyl scalar. 
By selecting appropriate bell-shaped or composite functions, we derive mass functions that generate geometries free of curvature singularities while satisfying asymptotic flatness and most energy conditions. 
These solutions not only demonstrate geometric regularity but also offer a broad range of physical behaviors depending on the model parameters.

An essential feature of our construction is the adoption of bell-shaped profiles, such as Gaussian, hyperbolic secant, and rational functions, for selected curvature invariants. 
This choice is motivated by both physical considerations and mathematical convenience. 
Physically, bell-shaped functions model the idea that quantum gravitational effects are localized near the black hole core and fade away rapidly at larger distances. 
Such behavior is consistent with the expectation that classical general relativity breaks down only in a finite high-curvature region, while remaining valid in the weak-field regime. 
In this sense, bell-shaped curvature profiles can be viewed as phenomenological representations of quantum corrections that regularize the spacetime at small scales. Mathematically, these functions are smooth, analytic, and decay sufficiently fast at infinity, thereby ensuring that the resulting spacetime is regular at the origin and asymptotically flat. 
Moreover, they allow for analytic control over the resulting geometry and often lead to effective energy-momentum tensors that respect standard energy conditions. 
By tuning the width and shape of these profiles, we can explore a wide range of regular spacetimes, including those mimicking classical black holes, those with de Sitter-like cores, and even traversable wormholes.

Our investigation into the QNMs of black holes constructed via the Weyl-scalar approach has revealed several critical insights into the interplay between effective potential structures and dynamical stability. 
Through a detailed analysis of the effective potentials and their corresponding time-domain profiles, we find that the height and width of the potential barrier govern the oscillation frequency and damping rate of QNMs in a predictable manner, with higher angular momentum modes exhibiting longer-lived and more oscillatory signals.


However, the presence of valleys in the potential profile introduces qualitatively new behavior.
We have shown that when the peak-to-valley ratio $|V_{\rm p}/V_{\rm v}|$ is large, the QNM waveform develops longer-lived oscillations, as perturbations are partially trapped in the valley before escaping.
In contrast, when the ratio is small, the valley is too shallow to induce significant trapping, and the waveforms reduce to standard exponential decay without sustained oscillatory features.
For example, the Sech-Weyl model with $\ell=2$ has a large ratio $|V_{\rm p}/V_{\rm v}| \approx 19.57$ and exhibits noticeably persistent oscillations, whereas the Gaussian-Weyl case with $\ell=2$ has a smaller ratio $|V_{\rm p}/V_{\rm v}| \approx 5.17$ and shows rapid, featureless decay.

These results emphasize that potential instabilities are not merely a consequence of asymmetry in the effective potential, but are fundamentally linked to the relative scale of the potential peak and valley. Our findings offer a clear diagnostic criterion for assessing the stability of black holes under linear perturbations and highlight the importance of fine potential features, especially in regular or modified gravity black holes.

Looking forward, this finite curvature approach can be extended in several promising directions. First, applying it to axisymmetric spacetimes may yield rotating regular black hole solutions that capture more realistic astrophysical features. Second, coupling this framework with modified gravity theories could provide insight into how higher curvature corrections influence both the geometry and QNM spectra. Finally, our results suggest that certain potential profiles may leave observable imprints in gravitational wave signals from compact object mergers. Future work may explore these phenomenological implications and confront them with data from current and upcoming detectors.

\section*{Acknowledgement}

L.C. was supported in part by the National Natural Science Foundation of China under Grant No.\ 12175108, and also by Yantai University under Grant No.\ WL22B224.
Z.-X. Z is also supported by the Pilot Scheme of Talent Training in Basic Sciences (Boling Class of Physics, Nankai University), Ministry of Education.

\appendix

\section{Axial perturbation}
\label{app:axial}

We consider linear perturbations of the background spacetime~\cite{Nollert:1999ji, Berti:2009kk, Konoplya:2011qq, Chen:2019iuo},
\begin{equation}
    g_{\mu\nu} = \mathring{g}_{\mu\nu}+h_{\mu\nu},
\end{equation}
where $\mathring{g}_{\mu\nu}$ denotes the background metric and $h_{\mu\nu}$ represents a small perturbation.  
In our case, the background metric is spherically symmetric and takes the form
\begin{equation}
    \mathring{g}_{\mu\nu} = \mathrm{diag}\left\{-f(r),\, f^{-1}(r),\, \frac{r^2}{1-\chi^2},\, r^2(1-\chi^2)\right\},
\end{equation}
where we have introduced the variable $\chi = \cos\theta$ to simplify the angular dependence.

The perturbation $h_{\mu\nu}$ can be decomposed into an axial (odd-parity) part $h^{(\rm{axial})}_{\mu\nu}$ and a polar (even-parity) part $h^{(\rm{polar})}_{\mu\nu}$. In this work we focus on the axial sector, whose non-vanishing components are
\begin{equation}
    h^{(\rm{axial})}_{03} = h^{(\rm{axial})}_{30}=-\epsilon \left(1-\chi ^2\right)  h_0(r)\, e^{- i \omega t} \, P'(\chi),
\end{equation}
\begin{equation}
    h^{(\rm{axial})}_{13} = h^{(\rm{axial})}_{31}=-\epsilon \left(1-\chi ^2\right)  h_1(r)\, e^{- i \omega t} \, P'(\chi),
\end{equation}
where $P(\chi)$ denotes the Legendre polynomial satisfying
\begin{equation}
  (1-\chi ^2) P''(\chi)-2 \chi P'(\chi)+\ell(\ell+1) P(\chi)=0.
\end{equation}
Here $\epsilon$ is a bookkeeping parameter that controls the perturbative expansion and keeps track of the order of the perturbation.

Then, we consider that the regular black hole is regarded as being sourced by an anisotropic fluid.
And the corresponding energy-momentum tensor for an anisotropic fluid is expressed as
\begin{equation}\label{eq:TAF}
    T_{\mu\nu} = (\rho+p_1)u_\mu u_\nu +(\rho-p_1)x_\mu x_\nu + p_1 g_{\mu\nu},
\end{equation}
where $\rho$ and $p_1$ are the energy density and tangential pressure, respectively. 
Here, $u^\mu$ is the unit timelike four-velocity, and $x^\mu$ is the unit spacelike vector that is always orthogonal to both $u^\mu$ and the angular directions. 
Therefore, $u^\mu$ and $x^\mu$ satisfy the constraints
\begin{equation}\label{eq:umu-xmu}
    u_\mu u^\mu = -1,\quad x_\mu x^\mu = 1,\quad u_\mu x^\mu = 0,
\end{equation}
and can be parameterized as
\begin{equation}\label{eq:umu-xmu-2}
    u_\mu = \left(u_t, 0, 0, 0\right), \quad x_\mu = \left(0, x_r, 0, 0\right).
\end{equation}

According to the Einstein field equation $R_{\mu\nu} - \frac{1}{2}g_{\mu\nu}R = 8\pi T_{\mu\nu}$, we can obtain 
\begin{equation} \label{eq:R}
R=-16\pi p_1. 
\end{equation}
The change of the energy-momentum tensor before and after the perturbation is
\begin{equation}
    \delta T_{\mu\nu} = \delta(\rho+p_1)u_\mu u_\nu+(\rho+p_1)\delta (u_\mu u_\nu) + \delta(\rho-p_1)x_\mu x_\nu + (\rho-p_1)\delta(x_\mu x_\nu) + \delta(p_1 g_{\mu\nu}).
\end{equation}
Enforcing the kinematic constraints on $x_\mu$ and $u_\mu$, we obtain
\begin{equation}
    \delta T_{13}=p_1  h^{(\rm{axial})}_{13}, \quad \delta T_{23}=0.
\end{equation}
And the master equations of axial perturbations are determined by
\begin{equation}\label{eq:delta-G}
    \delta G_{13}=\delta R_{13}-\frac{1}{2}Rh^{(\rm{axial})}_{13}= 8\pi \delta T_{13}, \delta G_{23}=\delta R_{23}= 8\pi \delta T_{23}.
\end{equation}
Therefore, according to Eq.~\eqref{eq:R} and~\eqref{eq:delta-G}, we can write the master equations of axial perturbation as
\begin{equation}
    \delta R_{13}=0,\quad \delta R_{23}=0
\end{equation}
which allow us to eliminate $h_0(r)$ and reduce the dynamics to a second-order equation for $h_1(r)$
\begin{multline}
    r^2 f(r)^2 h_1''(r)+r f(r)\big[3 r f'(r)-2 f(r)\big] h_1'(r) \\
    +h_1(r)\Big\{ f(r)\big[r^2 f''(r)+2 f(r)-\ell(\ell+1)\big]
    +r^2\big[f'(r)^2+\omega^2\big]\Big\}=0.
\end{multline}
Introducing the redefinition
\begin{equation}
    h_1(r) = \frac{r}{f(r)}\,\Psi(r),
\end{equation}
the equation becomes
\begin{equation}
    -r^2 f(r)^2 \Psi''(r)-r^2 f(r) f'(r)\Psi'(r)
    +\Psi(r)\left[f(r)\big(-3 r f'(r)+\ell(\ell+1)\big)-r^2\omega^2\right]=0.
\end{equation}
Next, by switching to the tortoise coordinate $r^*$ defined via
\begin{equation}
    \frac{d r^*}{d r} = \frac{1}{f(r)},
\end{equation}
the above equation can be cast into a Schrödinger-like form, known as the master equation for axial perturbations,
\begin{equation}
    -\frac{d^2\Psi}{d r^{*2}} + V_{\rm eff}(r)\,\Psi(r^*) = \omega^2 \Psi(r^*),
\end{equation}
with the effective potential given by
\begin{equation}
\label{eq:potential}
   V_{\rm eff}(r) = f(r)\left[\frac{\ell(\ell+1)}{r^2}-\frac{3 f'(r)}{r}\right].
\end{equation}



It should be noted that although this effective potential is formally consistent with the vacuum Schwarzschild black hole, it has already taken into account the contribution of the fluid term.
This result follows from the constraint relations imposed by the normalization of the four-velocity and the coordinate conditions, which eliminate independent perturbative degrees of freedom of the fluid sector at linear order. 
As a consequence, the effective potential Eq.\ \eqref{eq:potential} remains unaffected by explicit matter perturbations in this interpretation.

By contrast, if one assumes that regular black holes are generated by nonlinear electrodynamics, the situation changes substantially. In that case, the perturbation of the energy–momentum tensor necessarily contributes to the perturbation equations, leading to a coupled system of two non-decouplable equations. We adopt the first two schemes in this work. Addressing the problem of non-decoupled equations is the subject of another forthcoming work by us \cite{Yang:2026jch}.

\bibliographystyle{utphys}
\bibliography{references}
\end{document}